\documentclass[11pt,a4paper]{article}

\usepackage{colortbl}
\usepackage{ascmac}
\usepackage{bm}
\usepackage[dvipdfmx]{graphicx}
\usepackage{amsmath}
\usepackage{amsfonts}

\usepackage{tikz}
\usetikzlibrary{shapes}
\usetikzlibrary{arrows.meta}

\def\tr{\mathop{\rm tr}\nolimits}
\def\rank{\mathop{\rm rank}\nolimits}
\def\pexp{\mathop{\rm PE}\nolimits}

\newcommand{\wt}{\widetilde}

\newcommand{\ZZ}{\mathbb{Z}}
\newcommand{\RR}{\mathbb{R}}
\newcommand{\ol}{\overline}

\usepackage{subcaption}

\newcommand{\rap}[2]
{\setbox1=\hbox{#1}%
\setbox2=\hbox to\wd1{\hss #2\hss}%
\mbox{\rlap{\box1}\box2}}

\newcommand{\sla}[1]{\rap{$#1$}{$\backslash$}}

\newcommand{\ul}{\underline}

\begin{document}

\def\Pexp{\mathop{\rm PE}\nolimits}
\newcommand{\diff}{\mathrm{d}}
\def\newA{\notag \\ &\qquad}
\def\Total{\mathop{\rm Total}}


\begin{titlepage}
\title{
\vspace{-1.5cm}
\begin{flushright}
{\normalsize TIT/HEP-711\\  July 2026}
\end{flushright}
\vspace{1.5cm}
\LARGE{Line operator indices of S-fold theories}}
\author{
Yosuke {\scshape Imamura\footnote{E-mail: imamura@phys.sci.isct.ac.jp}},
Masato {\scshape Inoue\footnote{E-mail: masatoinoue@th.phys.titech.ac.jp}}, and
Akihiro {\scshape Sei \footnote{E-mail: a.sei@th.phys.titech.ac.jp}} \\
\\
\\
{\itshape Department of Physics, Institute of Science Tokyo}, \\ {\itshape Tokyo 152-8551, Japan}}

\date{}
\maketitle
\thispagestyle{empty}

\begin{abstract}
We study superconformal indices in the presence of line operators in S-fold theories.
We consider two types of lines,
realized by fundamental strings and fivebranes, respectively, in AdS$_5 \times S^5/\mathbb{Z}_k$.
For fundamental lines, we analyze the corresponding indices including
finite-$N$ corrections arising from giant graviton configurations.
For $k=2$, we compare the holographic results with line operator indices in $\mathcal{N}=4$
super Yang-Mills theories with orthogonal gauge groups.
For $k=3,4$ and $6$, we focus on the rank $2$ case, in which
the supersymmetry is enhanced to ${\cal N}=4$,
and compare them with the corresponding Wilson-'t Hooft line operator indices.
In both cases,
we find improved agreement once giant graviton contributions are included.
For line operators realized by fivebranes,
after confirming agreement for $k=2$,
we construct BPS configurations for $k\geq 3$ explicitly using fivebrane junctions
and derive the indices in the large $N$ limit by the mode analysis
on the fivebrane junctions.
\end{abstract}

\end{titlepage}

\tableofcontents

\section{Introduction}

In recent developments in quantum field theory,
non-Lagrangian field theories have played
an important role.
Among such theories, S-fold theories are
four-dimensional $\mathcal{N}=3$ superconformal field theories,
realized as generalized orbifolds involving S-duality
\cite{Garcia-Etxebarria:2015wns,Aharony:2015oyb,Aharony:2016kai}.
The presence of an $SL(2,\ZZ)$ duality twist
that mixes electric and magnetic charges
makes perturbative analysis difficult.
These theories lack known Lagrangian descriptions
except for exceptional cases,
yet they possess rich symmetry structures
and provide a useful setting for testing ideas in holography.

The main purpose of this paper is to study
BPS line operators in S-fold theories.
Line operators provide an important probe of the global structure
of quantum field theories \cite{Kapustin:2005py,Gaiotto:2014kfa}.
In particular, they are sensitive to generalized global symmetries
such as $1$-form symmetries \cite{Gaiotto:2014kfa,Aharony:2013hda}, 
and encode information that is not accessible through
local operators alone. 
In strongly coupled theories without a Lagrangian description, 
such as S-fold theories, it is in general difficult
to study line operators directly.

The AdS/CFT correspondence \cite{Maldacena:1997re,Gubser:1998bc,Witten:1998qj}
often provides
powerful tools for analyzing such strongly coupled theories.
In particular, S-fold theories
are defined by the brane construction, and
the AdS/CFT correspondence provides an efficient framework
for investigating them.

S-fold theories are classified by complex reflection groups \cite{Aharony:2016kai}.
We denote the theory associated with the complex reflection group
$G(k,k/\ell,N)$ by
$S_{k,\ell}(N)$.
$k=2,3,4,6$ is the order of the
orbifolding group $\ZZ_k$,
and $\ell$ is a divisor of $k$.
The holographic dual of $S_{k,\ell}(N)$ is
type IIB string theory in $AdS_5\times S^5/\ZZ_k$.
Line operators in S-fold theories were studied in
\cite{Etheredge:2023ler}
as a generalization of the $k=2$ case
studied in \cite{Bergman:2022otk} from the holographic viewpoint.
In \cite{Etheredge:2023ler,Bergman:2022otk} line operators were classified using the $1$-form symmetry
associated with two-form fields in $AdS_5$.
We consider two types of line operators.
The first class consists of fundamental-type line operators
realized by $(p,q)$-strings in $AdS_5$.
They generalize Wilson lines in the fundamental representation
of ${\cal N}=4$ SYM,
which are realized by fundamental strings \cite{Maldacena:1998im,Rey:1998ik}.
The second class consists of ``fat string'' line operators
realized by
$(p,q)$-fivebranes wrapped on four-cycles in $S^5/\ZZ_k$.
They generalize Wilson lines in spinor representations in $SO(N)$ SYM,
which are realized by
D5-branes wrapped on a four-cycle in the internal space \cite{Witten:1998xy}.
These two classes are classified by the following twisted homology groups of the internal space
$S^5/\ZZ_k$:
\begin{align}
\Gamma_k^{(1)}&=H^0(S^5/\ZZ_k,\wt{\ZZ^{F1}\times\ZZ^{D1}}),\nonumber\\
\Gamma_k^{(5)}&=H^4(S^5/\ZZ_k,\wt{\ZZ^{D5}\times\ZZ^{NS5}}),
\label{gk1gk5}
\end{align}
where the tildes indicate the $\ZZ_k$ twist.
$\Gamma_k^{(1)}$ and $\Gamma^{(5)}_k$ are isomorphic, and
$\Gamma_k^{(1/5)}$ for $k=2,3,4,6$ are \cite{Etheredge:2023ler}
\begin{align}
\Gamma_2^{(1/5)}=\ZZ_2\times\ZZ_2,\quad
\Gamma_3^{(1/5)}=\ZZ_3,\quad
\Gamma_4^{(1/5)}=\ZZ_2,\quad
\Gamma_6^{(1/5)}=0.
\end{align}

The classification of line operators based on 1-form symmetries
provides important topological information about these theories
\cite{Gaiotto:2014kfa,Aharony:2013hda}.
However, it does not capture full dynamical information associated with
these operators.

Supersymmetric observables
such as the superconformal index \cite{Romelsberger:2005eg,Kinney:2005ej}
offer a useful probe of
such dynamics and provide nontrivial tests of holography.
In this paper, we calculate superconformal indices
with the line-operator insertion \cite{Gomis:2011pf,Ito:2011ea,Gang:2012yr,Dimofte:2011py,Drukker:2015spa,Hatsuda:2023iwi,Guo:2023mkn,Hatsuda:2023imp,Hatsuda:2023iof}.

Studies of the superconformal index of S-fold theories
on the gravity side of the AdS/CFT correspondence
have been carried out for both
large $N$ \cite{Imamura:2016abe}
and finite $N$ \cite{Arai:2019xmp}.
Among various cases, the $N=2$ theories are of particular interest
because of the enhancement of the supersymmetry from ${\cal N}=3$ to ${\cal N}=4$ \cite{Aharony:2016kai}.
In this work, we extend the previous analysis to
the index with line-operator insertions.

This paper is organized as follows.
In the remainder of this section we define
the index and fix our conventions
so that it is consistent with the
supersymmetry preserved by $S$-folding and the
line operator insertions.
In Section \ref{fund.sec}, we study line operators realized by fundamental strings
and analyze their superconformal indices,
including finite-$N$ corrections from giant gravitons \cite{McGreevy:2000cw,Mikhailov:2000ya}.
We calculate the leading finite $N$ corrections
by singly wrapping giant gravitons.
We especially focus on the $N=2$ case,
and compare the results with those expected from the
supersymmetry enhancement.
In Section \ref{fat.sec}, we investigate fat-string type line operators
realized by fivebrane junctions
and derive their large-$N$ indices from fluctuation modes on them.
Section \ref{summary.sec} contains a summary and discussion.
Several technical details are collected in the appendices.

\subsection{Superconformal index and Schur index}
We will investigate the superconformal indices of S-fold theories $S_{k,1}$ ($k=2,3,4,6$) with
line operator insertions.
We need to carefully define the index so that it is compatible with the S-fold projection
and the line operator insertions.
We first summarize our conventions and notation.

Let
$Q_\pm^I$ and $\ol Q^{\dot\pm}_I$ ($I=1,2,3,4$)
be the supercharges of ${\cal N}=4$ SYM.
We choose two supercharges ${\cal Q}=Q_-^4$ and $\ol{\cal Q}=\ol Q^{\dot+}_1$
to define the superconformal index and its Schur limit.
The corresponding BPS bounds are
\begin{align}
\Delta&=\{{\cal Q},{\cal Q}^\dagger\}
=H-J_1+J_2-R_x-R_y+R_z\geq0,\nonumber\\
\ol\Delta&=\{\ol{\cal Q},\ol{\cal Q}^\dagger\}
=H-J_1-J_2-R_x-R_y-R_z\geq0.
\label{twobounds}
\end{align}
See Appendix \ref{killing.app} for the conventions for supercharges
and Cartan generators of the superconformal algebra
appearing in (\ref{twobounds}).
Although we can define the superconformal index
associated with $\ol{\cal Q}$ in the absence of line operators,
in the presence of line operators, $\ol{\cal Q}$ is not preserved,
and we cannot define the superconformal index.
We consider the insertion of line operators that break $J_2$ and $R_z$,
appearing in $\ol\Delta$.
This implies that the insertion of such line operators breaks $\ol{\cal Q}$.
Instead of the superconformal index, we use the Schur index \cite{Gadde:2011uv}
\begin{align}
I=\tr\left[(-1)^Fq^{J_1}x^{R_x}y^{R_y}\right],\quad
q=xy
\label{schurindex}
\end{align}
associated with $q^4_-={\cal Q}-\ol{\cal Q}$,
which is preserved
by BPS line operators
we studied in this paper.
Only operators saturating both the two bounds in (\ref{twobounds}) at the same time
contribute to the Schur index.
See \cite{Bourdier:2015wda,Pan:2021mrw,Hatsuda:2022xdv} for analytic expressions for
the Schur index of ${\cal N}=4$ $U(N)$ SYM without line insertion.

We define the S-folds using the
$\ZZ_k$ symmetry generated by
\begin{align}
U_k=\exp\left(\frac{2\pi i}{k}S\right),\quad
S=-R_x+R_y+R_z+A,
\label{zkgenerator}
\end{align}
where $A$ is the generator of the $U(1)$ R-symmetry of type IIB supergravity,
which we call $U(1)_A$ symmetry.
The $U(1)_A$ symmetry is broken to its discrete subgroup
by the flux and charge quantization.
For $k=1,2$, all supercharges are preserved by the $\ZZ_k$ projection, while
for $k\geq 3$, $Q_\pm^2$ and $\ol Q^{\dot\pm}_2$ are projected out
and ${\cal N}=3$ supersymmetric theory is realized.
Because both ${\cal Q}$ and $\ol{\cal Q}$ are preserved by the S-fold projection,
we can use the Schur index (\ref{schurindex}) to study
line operators in S-fold theories.

Before S-folding, line operators are $1/2$ BPS and preserve $8$ out of $16$ supersymmetries.
The S-folding with $k\geq3$ projects out four preserved supercharges,
and then the resulting configuration is $1/3$ BPS.
See Appendix \ref{killing.app}.

\section{Fundamental lines}\label{fund.sec}
In this section we study line operators realized by fundamental
strings.
We first review the known $k=1$ case (${\cal N}=4$ $U(N)$ SYM)
and the giant graviton expansion.
We then generalize the analysis to S-fold theories with $k=2$
(${\cal N}=4$ $SO(2N)$ SYM) and $k\geq 3$.
In the case of $k\geq3$,
the $N=2$ case is particularly
interesting because of the supersymmetry enhancement \cite{Aharony:2016kai}.
One of our goals is to investigate whether the line-operator index
is consistent with this proposal.

\subsection{$k=1$ case}
We briefly review the $k=1$ case
(${\cal N}=4$ $U(N)$ SYM) in this subsection
mainly to establish notation and to
illustrate the giant graviton expansion that will be generalized
to S-fold theories below.
In the case of the $k=1$ S-fold theory $S_{1,1}(N)$
the line operator index in the large $N$ limit was studied in \cite{Gang:2012yr}.
See also \cite{Hatsuda:2023imp,Hatsuda:2023iof}.
Fluctuation modes on a fundamental string in $AdS_2\subset AdS_5$ were studied in \cite{Drukker:2000ep,Faraggi:2011bb},
and 
it was confirmed that
the corresponding index correctly reproduces
the line operator index.

For finite $N$, the index acquires the contribution from giant gravitons
\cite{McGreevy:2000cw,Mikhailov:2000ya}, which are D3-branes
wrapped on three-cycles in the internal space. They represent
finite-$N$ effects beyond the supergravity approximation and
give rise to corrections suppressed by powers of $x^N$ and $y^N$. The resulting giant
graviton expansion provides a systematic organization of finite-$N$
corrections to the line-operator index.
See \cite{Imamura:2021ytr,Gaiotto:2021xce,Murthy:2022ien}
 for the giant graviton expansion for ${\cal N}=4$ $U(N)$ SYM without line operator insertion.
The giant graviton expansion of the index with fundamental line operators in ${\cal N}=4$ $U(N)$ SYM
was studied in \cite{Imamura:2024lkw,Beccaria:2024lbt}.
See also \cite{Imamura:2024zvw} for a generalization to line operators realized by multiple fundamental strings.

Let $I_{S_{1,1}(N),{\rm F1}}$ be the line operator index of $S_{1,1}(N)$
(${\cal N}=4$ $U(N)$ SYM)
with a line operator realized by a fundamental string.
It takes the form
\begin{align}
I_{S_{1,1}(N),{\rm F1}}=I_{\rm sugra}\sum_{m_x,m_y=0}^\infty
\sum_{\alpha=\bullet,x,y}x^{m_x N}y^{m_y N}
I_{\rm F1}^{(\alpha)}F^{(\alpha)}_{m_x,m_y},
\label{unfund}
\end{align}
where $I_{\rm sugra}$, $I_{\rm F1}^{(\alpha)}$, and $F_{m_x,m_y}^{(\alpha)}$
are the contributions from the supergravity multiplet in the bulk,
the fundamental string, and giant gravitons, respectively.
$m_x$ and $m_y$ are the wrapping numbers of D3-branes around the three-cycles $S_X$ and $S_Y$, respectively.
\footnote{Let us represent $S^5$ as $|X|^2+|Y|^2+|Z|^2=1$ with complex coordinates $X$, $Y$, and $Z$.
Then, $S_X$ and $S_Y$ are given by $X=0$ and $Y=0$, respectively.}
The additional label $\alpha=\bullet,x,y$ specifies the behavior of the string worldsheet
at the intersection with the giant gravitons;
for $\alpha=\bullet$ the infinite string does not end on the
giant gravitons, while for $\alpha=x$ and $\alpha=y$
the string is divided into two parts by the intersection,
and the two semi-infinite strings end on D3-branes wrapped on $S_X$ or $S_Y$.
Because the gauge fugacity integrals associated with the two cycles factorize \cite{Arai:2020qaj},
if two semi-infinite strings end on different cycles the integral vanishes
and such configurations
do not contribute to the index.
The contributions from the fundamental string are given by 
\cite{Imamura:2024lkw,Beccaria:2024lbt}
\begin{align}
I^{(\bullet)}_{\rm F1}&=\pexp(x+y-q),\nonumber\\
I_{\rm F1}^{(x)}&=x^{-1}\pexp(2y-2q),\nonumber\\
I_{\rm F1}^{(y)}&=y^{-1}\pexp(2x-2q),
\label{if1}
\end{align}
based on the mode analysis in
\cite{Drukker:2000ep,Faraggi:2011bb}.

$F_{m_x,m_y}^{(\alpha)}$ is the index of the gauge theory
realized on the giant gravitons.
For $\alpha=x,y$, the string endpoints on the giant gravitons play the role of line operators, and
$F_{m_x,m_y}^{(\alpha)}$ are the corresponding line operator indices.
$F_{m_x,0}^{(y)}=F_{0,m_y}^{(x)}=0$ by definition.
(The semi-infinite strings do not have giant gravitons to end on.)

Let us generalize
(\ref{unfund}) to $S$-fold theories $S_{k,1}$.
The form of the expansion (\ref{unfund}) does not change,
but we need to replace $I_{\rm sugra}$ and $F_{m_x,m_y}^{(\alpha)}$ appropriately as we will
explain below.

\subsection{Letter indices}

The fundamental string corresponding to a fundamental line operator is
localized at a point in the internal space $\bm{S}^5/\ZZ_k$.
$\ZZ_k$ projection identifies the string
and $k-1$ mirror images.
The independent degrees of freedom are
the same as those before the S-fold projection.
Hence,
the contribution from the fundamental string is the same as
the $k=1$ case (\ref{if1}).

On the other hand, modes of the supergravity multiplet in the bulk and
the vector multiplets on the giant gravitons are projected by $\ZZ_k$,
and we need to calculate the corresponding letter indices.
For that purpose it is convenient to define the
$U(1)_A$-refined index \cite{Arai:2019xmp}
\begin{align}
I=\tr[(-1)^Fq^{J_1}x^{R_x}y^{R_y}\eta^{R_z+A}],\quad
q=xy.
\label{refschur}
\end{align}
Due to the explicit symmetry breaking $U(1)_A\rightarrow\ZZ_k$,
the $U(1)_A$ fugacity $\eta$ takes only special values $\eta=e^{\frac{2\pi i}{k}r}$ ($r=0,\ldots,k-1$)
depending on $k$.
We can pick up
the $\ZZ_k$ invariant part from the refined indices by applying
the $\ZZ_k$ projection operator ${\cal P}_k$ defined by
\begin{align}
{\cal P}_kf(x,y,\eta)
=\frac{1}{k}\sum_{r=0}^{k-1}f(
e^{-\frac{2\pi i}{k}r}x,
e^{\frac{2\pi i}{k}r}y,
e^{\frac{2\pi i}{k}r}).
\end{align}

For example, the supergravity index in the S-fold background is
\begin{align}
I_{\rm sugra}(x,y)=\pexp({\cal P}_kf_{\rm sugra}(x,y,\eta)),
\label{Isugra}
\end{align}
where the refined letter index of the supergravity multiplet is
\cite{Arai:2019xmp}
\begin{align}
f_{\rm sugra}&=\frac{(1-\eta^{-1}q)(1-\eta q)}{(1-x)(1-y)(1-q)}-\frac{1}{1-q}.
\label{fsugra}
\end{align}
This is obtained by summing up the contributions of all modes in $AdS_5\times S^5$ obtained in
\cite{Gunaydin:1984fk,Kim:1985ez}.

The index of the vector multiplet on a single giant graviton is obtained by
\begin{align}
F_{1,0}=\pexp\left({\cal P}_k\sigma_xf_{\rm vec}\right),\quad
F_{0,1}=\pexp\left({\cal P}_k\sigma_yf_{\rm vec}\right),\quad
\end{align}
where
\begin{align}
f_{\rm vec}&=\eta\left(1-\frac{(1-\eta^{-1}x)(1-\eta^{-1}y)}{1-q}\right)
\label{fvecrefined}
\end{align}
is the refined letter index of the vector multiplet,
and $\sigma_x$ and $\sigma_y$ are
involution relating the symmetry generators
acting on the boundary and those on the giant gravitons \cite{Arai:2019xmp,Gaiotto:2021xce}
\begin{align}
\sigma_x :(q,x,y,\eta)\rightarrow(y,x^{-1},q,\eta^{-1}),\quad
\sigma_y :(q,x,y,\eta)\rightarrow(x,q,y^{-1},\eta^{-1}).
\label{sxsy}
\end{align}

In the derivation of (\ref{fvecrefined}),
we define the $U(1)_A$ action on the component fields of the vector multiplet
so that the scalar fields are $U(1)_A$ invariant.
If $U(1)_A$ is a continuous symmetry,
this must be the case because the scalar fields are real.
For a discrete subgroup $\ZZ_k$ with even $k$,
we can introduce additional non-trivial intrinsic parity.
In the $k=2$ case, this ambiguity enables us
to define two types of orientifolds
with $so$ and $sp$ gauge algebras.

\subsection{$k=2$ case}

Let us discuss the $k=2$ case in detail.
Because $S_{2,1}(N)$ is the orientifold theory with $G=SO(2N)$,
we can treat both the boundary theory and the bulk theory perturbatively,
and we can make a comparison between the results obtained on the two sides of the
AdS/CFT correspondence.

In this subsection we treat the index as a $y$-series expansion.
Namely, we expand the index in the form
\begin{align}
I(x,y)=\sum_{n=0}^\infty y^nf_n(x).
\label{yexp}
\end{align}
It is known that in many cases
only one cycle contributes to the index
in the form (\ref{yexp}),
and the giant graviton expansion reduces to a
simple-sum expansion \cite{Gaiotto:2021xce,Imamura:2022aua,Fujiwara:2023bdc}.

The simple-sum giant graviton expansion for
orientifold theories without line-operator insertions
was studied in \cite{Fujiwara:2023bdc}.
The purpose of this subsection is to generalize it
to the index with the fundamental line insertion.

The simple-sum version of (\ref{unfund}) is
\begin{align}
I_{S_{2,1}(N),{\rm F1}}
&=I_{\rm sugra}
\sum_{m_x=0}^\infty
x^{m_xN}\left(I_{\text{F1}}^{(\bullet)}F_{m_x,0}^{(\bullet)}+I_{\text{F1}}^{(x)}F_{m_x,0}^{(x)}\right).
\label{adsside}
\end{align}
$I_{\rm sugra}$ is the letter index of the supergravity multiplet in $AdS_5\times S^5/\ZZ_2$
and is obtained by the $\ZZ_2$ projection (\ref{Isugra}) \cite{Imamura:2016abe}.
It is also derived on the gauge theory side in \cite{Hatsuda:2024lcc}.

On the AdS side, the index is obtained in the form
(\ref{adsside}),
and the contribution from the fundamental string is the same as (\ref{if1}) in the $k=1$ case,
while $I_{\rm sugra}$ and $F_{m_x,0}^{(\alpha)}$ must be replaced appropriately.
$F^{(\alpha)}_{m_x,0}$ in the $k=2$ case is the index
of the gauge theory realized on the stack
of $m_x$ giant gravitons
wrapped around $S^3/\ZZ_2$.
Although the gauge group of the theory
is $U(m_x)$ locally,
it is broken to $O(m_x)$ by
the twist around the non-trivial cycle in $S^3/\ZZ_2$.
The twist acts on the gauge group as
an outer automorphism that swaps the fundamental and anti-fundamental representations.
This divides the adjoint representation of $U(m_x)$
into two parts,
the symmetric and the anti-symmetric representations of $O(m_x)$, with different $\ZZ_2$ parity.
Correspondingly, we define the $\ZZ_2$ refined adjoint character
\begin{align}
\chi_{\rm adj}^{U(m)}(U,\eta)
=\chi_{\rm sym}^{O(m)}(U)
+\eta\chi_{\rm adj}^{O(m)}(U).
\end{align}
We assign negative parity to $\chi_{\rm adj}^{O(m)}$.
This is consistent with the definition of the refined
letter index (\ref{fvecrefined}),
in which we assign negative parity to the gauge field.
The letter index $i_{\rm GG}$ of the fields on the $m_x$ coincident giant gravitons
is
\begin{align}
i_{\rm GG}={\cal P}_2(\sigma_xf_{\rm vec}\chi_{\rm adj}^{U(m_x)}(U)),
\end{align}
and $F_{m_x,0}^{(\alpha)}$ is given by
\begin{align}
F_{m_x,0}^{(\bullet)}&=\int_{O(m_x)} dU(\pexp i_{\rm GG}),\nonumber\\
F_{m_x,0}^{(x)}&=\int_{O(m_x)} dU(\pexp i_{\rm GG})|\chi_{\rm vect}(U)|^2.
\label{fm0z2}
\end{align}

Let us first discuss the half-BPS indices obtained from the Schur index by taking the
limit $y\rightarrow 0$ with $x$ fixed.
In this case, the $U(1)_A$-refined letter index
of the vector multiplet index is $f_{\rm vec}=x$,
and the projected letter index for the giant gravitons wrapped on the $S_X$
cycle is
\begin{align}
{\cal P}_2(\sigma_xf_{\rm vec}\chi_{\rm adj}^{U(m)})=x^{-1}\chi_{\rm adj}^{O(m)}
\end{align}
and this is identical with the involution of the letter index for the ${\cal N}=4$
$O(m)$ SYM defined in $S^3\times\RR$.
Namely,
the giant graviton indices
$F_{m_x,0}^{(\bullet)}$ and
$F_{m_x,0}^{(x)}$ are obtained from the
index of $O(N)$ SYM with and without vector line operator insertion
by applying $\sigma_x$:
\begin{align}
F_{m_x,0}^{(\bullet)}=\sigma_xI_{O(m)},\quad
F_{m_x,0}^{(x)}=\sigma_xI_{O(m),{\rm vector}}.
\end{align}

The half-BPS index is defined as the $y\rightarrow 0$ limit with $x$ fixed.
A lot of analytic results have been obtained in the literature.
The half BPS indices of ${\cal N}=4$ SYM with the gauge group $O(m)$
and $SO(2N)$ without line insertion are \cite{Hatsuda:2024lcc}
\begin{align}
I_{SO(2N)}&=\frac{1-x^{2N}}{1-x^N}I_{O(2N)},\quad
I_{O(m)}=\prod_{k=1}^{[\frac{m}{2}]}\frac{1}{1-x^{2k}},
\label{halfbps1}
\end{align}
where $[\cdots]$ denotes the floor function.
The half-BPS indices with line insertion are \cite{Hatsuda:2025jze}
\begin{align}
\frac{I_{SO(2N),{\rm vector}}}{I_{SO(2N)}}=\frac{(1+x^{N-1})(1-x^N)}{1-x},\quad
\frac{I_{O(m),{\rm vector}}}{I_{O(m)}}=\frac{1-x^m}{1-x}.
\label{halfbps2}
\end{align}
Using these analytic expressions, we can easily confirm the relation
(\ref{adsside})
holds in the half BPS limit $y\rightarrow0$.

\begin{figure}[htb]
\centering
\includegraphics[width=1\textwidth]{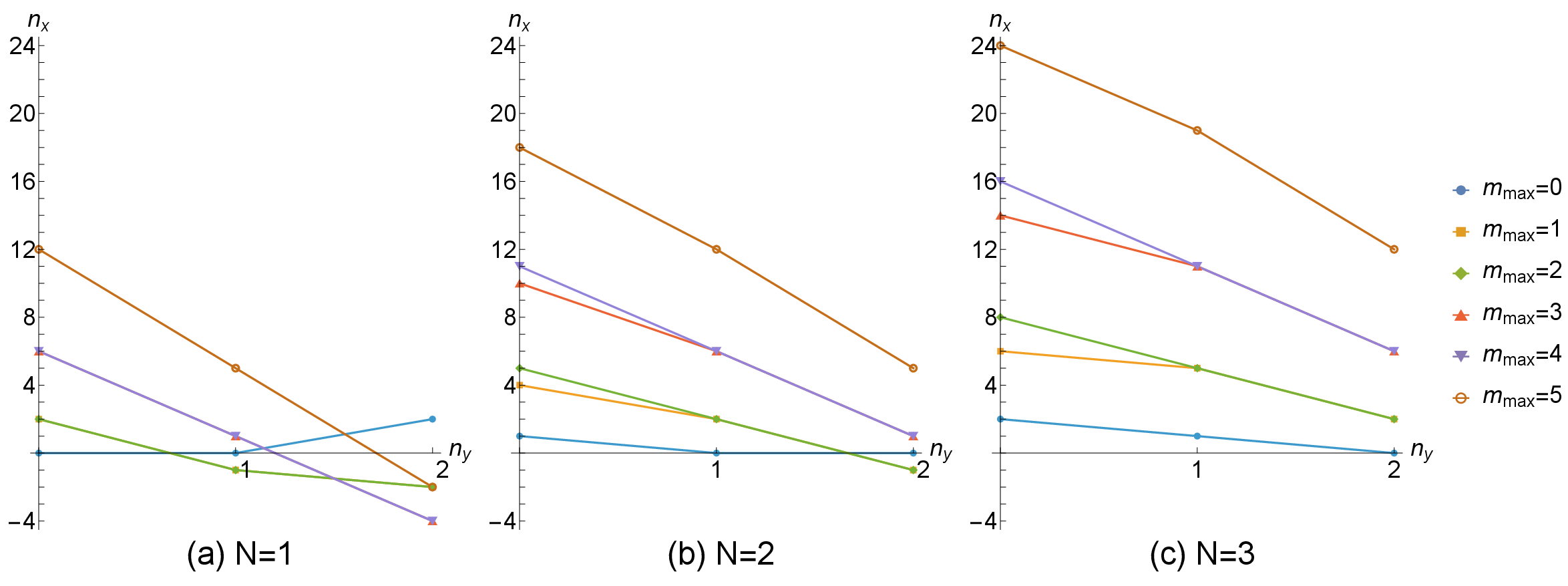}
\caption{The leading order in $x$ at each order of $y$ in the expansion of ${\Delta}_{SO(2N),\text{vec}}^{(m_{\text{max}})}$ up to $m_{\text{max}} \le 5$: 
$n_x$ is the leading order in $x$, and $n_y$ is the order in $y$.}\label{DeltaSO2Nvec}
\end{figure}

For the Schur index containing higher order terms in $y$,
it is difficult to analytically confirm (\ref{adsside}).
We use numerical analysis.
On the gravity side, we calculate
$I_{S_{2,1}(N),\text{F1}}^{(m \le m_{\text{max}})}$,
which includes the contribution from
giant gravitons with wrapping number $m \le m_{\text{max}}$,
and compare the results with the index
$I_{\rm SO(2N),{\rm vector}}$
calculated on the gauge theory side.
We expand the discrepancy to $y$-series as
\begin{align}
I_{S_{2,1},\text{F1}}^{(m \le m_{\text{max}})}-I_{\rm SO(2N),{\rm vector}}
=\sum_{n=0}^\infty c_n(x)y^n,
\label{discrepancy}
\end{align}
and expect the orders of the coefficients $c_n(x)$
to increase as we increase $m_{\rm max}$.
We show the result for $N=2$ in Figure \ref{DeltaSO2Nvec}.
We find that the orders increase for $N=2$ and $N=3$ as we increase $m_{\rm max}$.
For $N=1$,
one finds that the leading order occasionally shifts
to a lower order when
$m_{\rm max}$ is increased.
This behavior is nevertheless expected.
This is because the giant graviton index includes
negative power terms in $x$, so
the approximation may temporarily become worse for small $m$ and $N$.
See 
\cite{Gaiotto:2021xce,Imamura:2022aua,Fujiwara:2023bdc}
for some results of numerical analyses.
We expect the expansion converges to the
correct index in the $m_{\rm max}\rightarrow \infty$ limit.


\subsection{$k=3,4,6$ cases}
\subsubsection{Gravity side}
Let us consider the cases with $k=3,4$, and $6$.
S-fold theories with $k\geq 3$ are intrinsically strongly coupled,
and analysis in the boundary theory is in general difficult.
This is also the case for the multiply wrapped giant gravitons on the gravity side.
If $m_x+m_y\geq2$, the calculation with simple projection ${\cal P}_k$ does not work for $k\geq3$
because the $\ZZ_k$ action on the fields
on giant gravitons involves $SL(2,\ZZ)$ duality,
and it cannot be described by a local transformation
of elementary fields.
However, it is straightforward to
calculate the large $N$ limit index and the leading giant graviton
contribution with $m_x+m_y=1$.
For this reason,
we include only contributions with $m_x+m_y=0$ and $1$ in the following analysis.

We calculate the large $N$ index
\begin{align}
I_{S_{k,1}(N),{\rm F1}}^{(m=0)}=I_{\rm sugra}I_{\rm F1}^{(\bullet)}
\end{align}
and include the contribution from a single giant graviton
as the leading finite $N$ correction.
\begin{align}
I_{S_{k,1}(N),{\rm F1}}^{(m\leq1)}
&=
I_{S_{k,1}(N),{\rm F1}}^{(m=0)}+
I_{\rm sugra}\sum_{\alpha= \bullet ,x,y}I_{\rm F1}^{(\alpha)}(x^NF_{1,0}^{(\alpha)}+y^NF_{0,1}^{(\alpha)}).
\end{align}
The fundamental string contributions $I_{\rm F1}^{(\alpha)}$ are given by (\ref{if1}),
and the giant graviton indices are
\begin{align}
F_{1,0}^{(\bullet)}&=F_{1,0}^{(x)}=\pexp\left({\cal P}_k(\sigma_x f_{\rm vec})\right),\nonumber\\
F_{0,1}^{(\bullet)}&=F_{0,1}^{(y)}=\pexp\left({\cal P}_k(\sigma_y f_{\rm vec})\right).
\end{align}

It would be desirable to confirm the matching of these indices with the
results in the boundary theory.
However, this is difficult
for general $N$ because
no method is currently
known for calculating the indices for general $N$
in the boundary theories.
There is an exceptional case, $N=2$,
for which it was conjectured in \cite{Aharony:2016kai} that
rank $2$ S-fold theories $S_{k,1}(2)$
are equivalent to 
${\cal N}=4$ SYM with rank $2$
gauge groups.
For this reason, we focus on the $N=2$ case.

To show the results, we use variables
\begin{align}
t=\sqrt{xy},\quad
u=\sqrt{x/y},
\end{align}
and express them
as Taylor series expansions in the
variable $t$.
In the following, we call this the ``$t$-expansion''.
An advantage of the $t$-expansion
is that the giant graviton indices do not include negative powers of $t$,
and the convergence of the sum
over the wrapping numbers is obvious.

\subsubsection{$k=3$}

Before analyzing the line operator indices,
let us recall how the giant graviton expansion works
for the indices without line operator insertions \cite{Arai:2019xmp}.
The Schur index of the ${\cal N}=4$ SYM with $G=SU(3)$ is
\begin{align}
I_{SU(3)}&=1+0t+(u^2+1+\tfrac{1}{u^2})t^2
+(u^3-u-\tfrac{1}{u}+\tfrac{1}{u^3})t^3
+(u^4+2+\tfrac{1}{u^4})t^4
\nonumber\\&
+(u^5+u^3-2u-\tfrac{2}{u}+\tfrac{1}{u^3}+\tfrac{1}{u^5})t^5
+\cdots
\end{align}
On the gravity side, without including
giant graviton contributions,
we obtain
\begin{align}
I_{S_{3,1}(2)}^{(m=0)}&=1+0t+t^2+(u^3-u-\tfrac{1}{u}+\tfrac{1}{u^3})t^3
+(-u^2+4-\tfrac{1}{u^2})t^4
\nonumber\\&
+(3u^3-3u-\tfrac{3}{u}+\tfrac{3}{u^3})t^5+\cdots
\end{align}
We find agreement only for the first two terms in the
$t$ expansion.
We expect the inclusion of the giant graviton
contributions to improve the agreement.
The result including singly wrapped giant gravitons is
\begin{align}
I_{S_{3,1}(2)}^{(m\leq1)}&=1+0t+(u^2+1+\tfrac{1}{u^2})t^2
+(u^3-u-\tfrac{1}{u}+\tfrac{1}{u^3})t^3
+(u^4+2+\tfrac{1}{u^4})t^4
\nonumber\\&
+(u^7+u^3-2u-\tfrac{2}{u}+\tfrac{1}{u^3}+\tfrac{1}{u^7})t^5+\cdots
+\cdots.
\end{align}
Now, we find agreement for the first five terms,
and the inclusion of the giant graviton contributions
improves the result.
This provides strong evidence that the giant graviton expansion
remains valid for S-fold theories.
The coefficient of $t^5$ in $I_{S_{3,1}(2)}^{(m\leq1)}$
includes $u^7$ and $u^{-7}$, which correspond to
negative powers of $x$ and $y$.
Such terms often appear in giant graviton contributions
and should be canceled by contributions from
multiply wrapped configurations.

Let us turn to the line-operator index.
The supergravity index without giant-graviton contributions is
\begin{align}
I_{S_{3,1}(2),{\rm F1}}^{(m=0)}
&=1+(u+\tfrac{1}{u})t+(u^2+1+\tfrac{1}{u^2})t^2+(2u^3+\tfrac{2}{u^3})t^3+\cdots.
\label{is3f10}
\end{align}
By adding the singly wrapped giant graviton contributions,
we obtain
\begin{align}
I_{S_{3,1}(2),{\rm F1}}^{(m\leq 1)}
&=1+(2u+\tfrac{2}{u})t+(3u^2+3+\tfrac{3}{u^2})t^2
\nonumber\\&
+(u^5+4u^3+u+\tfrac{1}{u}+\tfrac{4}{u^3}+\tfrac{1}{u^5})t^3+\cdots
\label{is3f11}
\end{align}
We expect this to have better agreement with the corresponding line operator indices.
Because the coefficient of $t^3$
includes negative powers of $x$ and $y$,
agreement at this order cannot be expected.
However, it would be encouraging if we could
find better agreement for terms up to quadratic order.

We now ask which line operator
in the ${\cal N}=4$ SYM corresponds
to the fundamental string in AdS.
Up to screening, line operators of ${\cal N}=4$ SYM with $G=SU(N)$ are classified by
$\ZZ_N\times\ZZ_N$ charges.
The two $\ZZ_N$ factors
correspond to the electric and magnetic $N$-alities, respectively.
Let $\ell_{e,m}$ ($e,m=0,1,\ldots,N-1$) be the line operator
with charge $(e,m)$.
Consistency with the Dirac quantization condition implies
that a physically admissible set of line operators
must be mutually local.
Namely,
\begin{align}
em'-me'\in N\ZZ.
\end{align}
In the case with $N=3$, the admissible set of line operators is unique up to
$SL(2,\ZZ)$ duality,
and in an appropriate duality frame,
the only allowed line operators are Wilson line operators $\ell_{e,0}$,
labeled by $e=0,1,2$.

Furthermore, the line operators $\ell_{e,0}$ and $\ell_{-e,0}$ are
related by charge conjugation.
Therefore, in an appropriate duality frame,
we can always transform a line operator with
non-trivial charges into the fundamental
Wilson line operator $\ell_{1,0}$.
The line operator index with the fundamental Wilson line $\ell_{1,0}$ is
\begin{align}
I_{SU(3),{\rm fund}}
&=1+(u+\tfrac{1}{u})t+(2u^2+1+\tfrac{2}{u^2})t^2
+(2u^3+\tfrac{2}{u^3})t^3
\nonumber\\&
+(3u^4+\tfrac{3}{u^4})t^4+\cdots.
\label{su3fundline}
\end{align}
Disappointingly, the comparison of
this index with
(\ref{is3f10})
and (\ref{is3f11}) is contrary to
our expectations.
Although we find agreement up to the second term
in (\ref{is3f10}),
after including the singly wrapped giant graviton contributions,
the agreement becomes worse,
and (\ref{is3f11}) and
(\ref{su3fundline}) agree only in the first, trivial term.

This result seems to suggest that the identification
of the fundamental string with the fundamental Wilson line operator
is incorrect.
In the classification by the $1$-form symmetry,
two Wilson lines related by the insertion of
a dynamical local operator are identified.
Namely, the $1$-form symmetry classifies line operators
up to screening.
However, even if two line operators are in the same class,
the corresponding line operator indices are generally
different.
Namely, we need information finer than that encoded by the
$1$-form symmetry charges.

The line operator spectrum is
studied in \cite{Amariti:2023hev}
using $(p,q)$-string junctions.
See also \cite{Imamura:2016udl,Agarwal:2016rvx}.
A string junction can be specified by the charges $(p_i,q_i)$
of strings ending on D3-branes labelled by $i=1,\ldots,N$.
We are interested in the case with two D3-branes,
and then the charges are specified by a four-component vector
$(p_1,q_1;p_2,q_2)$.
For general string junctions that may have external strings,
the four charges $p_1$, $q_1$, $p_2$, and $q_2$ are independent
and span the lattice $\Lambda=\ZZ^4$.
However, junctions corresponding to dynamical particles
should not have external strings,
and such junctions form a sublattice $\Lambda_{\rm dyn}\subset\Lambda$.
The charges of line operators are classified by $\Lambda$.
The screening is effectively described by the identification $\Lambda_{\rm dyn}\sim 0$,
and the $1$-form symmetry charges of the line operators are classified by $\Lambda/\Lambda_{\rm dyn}$.

Following \cite{Amariti:2023hev}, let us take the following basis
\begin{align}
w_1&=(1,0;-1,0),&
w_2&=(0,1;1,1),\nonumber\\
m_1&=(0,1;0,-1),&
m_2&=(-1,-1;-1,0).
\label{w1w2m1m2}
\end{align}
The Dirac pairings among them are
\begin{align}
&\langle w_1,w_2\rangle=\langle w_1,m_2\rangle=0,\nonumber\\
&\langle w_1,m_1\rangle=\langle w_2,m_2\rangle=2,\nonumber\\
&\langle w_1,m_2\rangle=\langle w_2,m_1\rangle=-1.
\end{align}
These relations are the ones for the ${\cal N}=4$ $SU(3)$ SYM.
We can easily check that $\Lambda/\Lambda_{\rm dyn}=\ZZ_3$.
In an appropriate duality frame,
we can identify this with the $3$-ality of the Wilson line operators.

\begin{figure}[htb]
\centering
\includegraphics{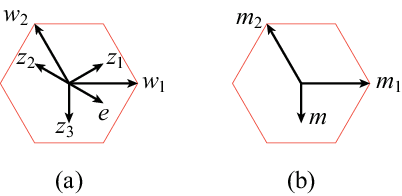}
\caption{The electric component $e$ and the magnetic component $m$
of the charge $q_{\rm F1}$ are shown as vectors on
the weight lattices of
the electric $su(3)$ and the magnetic $su(3)={}^Lsu(3)$.
The vectors labeled by $z_1$, $z_2$, and $z_3$ represent
the charges associated with the fugacities $z_1$, $z_2$, and $z_3=(z_1z_2)^{-1}$
in (\ref{su3chmche}).}\label{su3dyon.eps}
\end{figure}

We would like to identify the line operator
corresponding to a fundamental string
ending on one of the D3-branes.
Namely, we are interested in the line operator corresponding to the
vector
\begin{align}
q_{\rm F1}=(1,0;0,0).
\end{align}
One can easily expand this in terms of the four vectors in 
(\ref{w1w2m1m2}) as (Figure \ref{su3dyon.eps})
\begin{align}
q_{\rm F1}=\frac{1}{3}w_1-\frac{1}{3}w_2-\frac{1}{3}m_1-\frac{2}{3}m_2.
\label{f1charge}
\end{align}
If we allow screening, this would be in the same class as
\begin{align}
q_{\rm F1}\sim
q_{\rm F1}+m_2=\frac{1}{3}(w_1-m_1)-\frac{1}{3}(w_2-m_2),
\label{f1m2}
\end{align}
and by the $SL(2,\ZZ)$ duality transformation
\begin{align}
\left(\begin{array}{c}
w'_i \\
m_i'
\end{array}\right)
=
\left(\begin{array}{cc}
1 & -1  \\
0 & 1
\end{array}\right)
\left(\begin{array}{c}
w_i \\
m_i
\end{array}\right),
\end{align}
this would become
the fundamental line operator with charge
\begin{align}
q_{\rm F1}\sim
q_{\rm F1}+m_2=\frac{1}{3}w'_1-\frac{1}{3}w_2'.
\end{align}
However, the information that we need is not this.
As we mentioned above,
we should not use the equivalence `$\sim$' by the screening,
and then we cannot rewrite the
line operator with the charge
(\ref{f1charge}) as a Wilson-line operator.
We adopt the dyonic charge
(\ref{f1charge}) as it is,
and the corresponding line operator is
a 't Hooft-Wilson line operator.

The line-operator index for 't Hooft line operators
is investigated in \cite{Gang:2012yr}.
Let $m\in P^\vee$ be the GNO magnetic charge of the 't Hooft line.
In general this breaks the gauge symmetry algebra $\mathfrak{g}$ to $\mathfrak{g}_m$.
If $m$ is the highest weight of
a minuscule representation of ${}^L\mathfrak{g}$,
we do not need to take into account the monopole bubbling phenomenon
\cite{Kapustin:2006pk,Ito:2011ea,Brennan:2018yuj,Assel:2019iae}
and the index is given by \cite{Gomis:2011pf,Ito:2011ea,Gang:2012yr}
\begin{align}
I_{G,m}=\frac{1}{|W(\mathfrak{g}_m)|}
\int\prod_{i=1}^{\rank G}\frac{d\lambda_i}{2\pi}\pexp
\left[\rank G-\frac{(1-x)(1-y)}{1-q}\chi_m\right],
\label{thooftindex}
\end{align}
where $\chi_m$ is the adjoint character modified by the magnetic flux
\begin{align}
\chi_m=\sum_\alpha q^{\frac{|\alpha(m)|}{2}}
e^{i\alpha(\lambda)},
\end{align}
and $|W(\mathfrak{g}_m)|$ is the order of the Weyl group of the
unbroken gauge algebra $\mathfrak{g}_m$.
The integral is taken over the maximal torus $\mathfrak{h}/Q^\vee$,
where $\mathfrak{h}\subset\mathfrak{g}$ is the Cartan
subalgebra of $\mathfrak{g}$.

We want to generalize the formula (\ref{thooftindex}) to the
Wilson-'t Hooft line operators with dyonic charge $(e,m)$.
The electric charge $e$ can be regarded
as the weight vector associated with a $\mathfrak{g}_m$ irreducible representation
$R_e$.
Let $\chi_e$ be the $\mathfrak{g}_m$ character of the representation $R_e$.
We assume that (\ref{thooftindex}) can be generalized to dyonic lines
by simply inserting the factor $|\chi_e|^2$ in the integrand,
where $\chi_e$ is the character of $R_e$
specifying the electric charge of the dyonic line:
\begin{align}
I_{G,(m,e)}=\frac{1}{|W(\mathfrak{g}_m)|}
\int\prod_{i=1}^{\rank G}\frac{d\lambda_i}{2\pi}\pexp
\left[\rank G-\frac{(1-x)(1-y)}{1-q}\chi_m\right]|\chi_e|^2
\label{ibr}
\end{align}

For the line operator with charge (\ref{f1charge}),
the modified character for $m$ and the $\mathfrak{g}_m$ character for $e$
appearing in
(\ref{ibr})
are explicitly given by
\begin{align}
\chi_m=2+
\frac{z_1}{z_2}+\frac{z_2}{z_1}
+\left(\frac{z_3}{z_1}+\frac{z_3}{z_2}+\frac{z_1}{z_3}+\frac{z_2}{z_3}\right)q^{\frac{1}{2}},\quad
\chi_e=\frac{1}{z_1}+\frac{1}{z_2}.
\label{su3chmche}
\end{align}
Formula
(\ref{ibr}) with these characters gives
the line operator index
\begin{align}
I_{SU(3),(m,e)}
&=1+(2u+\tfrac{2}{u})t+(3u^2+3+\tfrac{3}{u^2})t^2+(4u^3+u+u^{-1}+\tfrac{4}{u^3})t^3+\cdots
\label{su3d}
\end{align}
Comparing this with (\ref{is3f11}), we find that
the first three terms, which are expected to
match, indeed agree exactly.
The result suggests that the giant graviton expansion works well
and our identification of the line operator is correct.

\subsubsection{$k=4$}
Next, let us consider the $S_{4,1}(2)$ theory,
which is dual to ${\cal N}=4$ SYM with $G=SO(5)$.

Let us first review the results
in the absence of the line operator insertion \cite{Arai:2019xmp}.
The supergravity index is
\begin{align}
I_{S_{4,1}(2)}^{(m=0)}
&=1+0t+t^2
+(-u-\tfrac{1}{u})t^3
+(u^4+4+\tfrac{1}{u^4})t^4
+\cdots
\end{align}
and the inclusion of the singly wrapped giant graviton contributions gives
\begin{align}
I_{S_{4,1}(2)}^{(m\leq1)}
&=1+0t+(u^2+1+\tfrac{1}{u^2})t^2
+(-2u-\tfrac{2}{u})t^3
\nonumber\\&
+(u^4+3u^2+4+\tfrac{3}{u^2}+\tfrac{1}{u^4})t^4
+\cdots.
\end{align}
The comparison to the gauge theory result
\begin{align}
I_{SO(5)}
&=1+0t+(u^2+1+\tfrac{1}{u^2})t^2+(-2u-\tfrac{2}{u})t^3
\nonumber\\&
+(2u^4+3u^2+5+\tfrac{3}{u^2}+\tfrac{2}{u^4})t^4
+\cdots
\end{align}
again shows that the inclusion of giant gravitons improves the
agreement,
although the improvement is not as good as in the $k=3$ case.

Let us turn to the insertion of line operators.
On the gravity side, the inclusion of the fundamental string contribution gives
\begin{align}
I_{S_{4,1}(2),{\rm F1}}^{(m=0)}
&=1+(u+\tfrac{1}{u})t+(u^2+1+\tfrac{1}{u^2})t^2
+(u^3+\tfrac{1}{u^3})t^3
\nonumber\\&
+(2u^4+2+\tfrac{2}{u^4})t^4+\cdots
\end{align}
for no giant graviton contribution,
and
\begin{align}
I_{S_{4,1}(2),{\rm F1}}^{(m\leq1)}
&=1+(2u+\tfrac{2}{u})t+(2u^2+3+\tfrac{2}{u^2})t^2
+(3u^3+u+\tfrac{1}{u}+\tfrac{3}{u^3})t^3
\nonumber\\&
+(4u^4+u^2+\tfrac{1}{u^2}+\tfrac{4}{u^4})t^4
+(u^7+5u^5+u^3+\tfrac{1}{u^3}+\tfrac{5}{u^5}+\tfrac{1}{u^7})t^5+\cdots
\end{align}
with singly wrapped giant gravitons included.

This result does not agree well with the simple Wilson line indices
of the vector and the spinor representations:
\begin{align}
I_{SO(5),{\rm vec}}
&=1+(u+\tfrac{1}{u})t+(2u^2+1+\tfrac{2}{u^2})t^2
+(2u^3+\tfrac{2}{u^3})t^3
\nonumber\\&
+(4u^4+2u^2+3+\tfrac{2}{u^2}+\tfrac{4}{u^4})t^4+\cdots,\nonumber\\
I_{SO(5),{\rm spin}}
&=1+(u+\tfrac{1}{u})t+(2u^2+1+\tfrac{2}{u^2})t^2
+(2u^3+\tfrac{2}{u^3})t^3
\nonumber\\&
+(3u^4+u^2+1+\tfrac{1}{u^2}+\tfrac{3}{u^4})t^4+\cdots.
\end{align}

\begin{figure}[htb]
\centering
\includegraphics{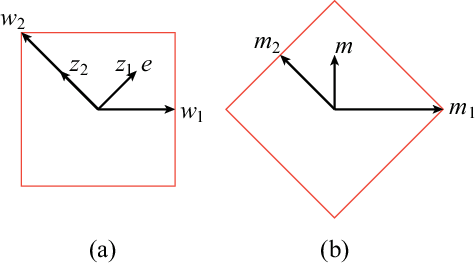}
\caption{The electric component $e$ and the magnetic component $m$ of the charge $q_{\rm F1}$ are
shown as vectors in the weight lattices of $so(5)$ and $usp(4)={}^Lso(5)$.
The vectors labeled by $z_1$ and $z_2$
represent charges associated with the fugacities used in (\ref{so5chmche}).}\label{so5dyon.eps}
\end{figure}

Let us determine the charges of the corresponding line operator as in the $k=3$ case.
Following \cite{Amariti:2023hev}, we use the following
charge vectors for dynamical particles.
\begin{align}
w_1&=(1,0;-1,0),&
w_2&=(-1,-1;1,-1),\nonumber\\
m_1&=(-1,1;1,-1),&
m_2&=(1,0;0,1).
\end{align}
The Dirac pairings of the corresponding line operators are
\begin{align}
&\langle w_1,w_2\rangle
=\langle m_1,m_2\rangle=0,\nonumber\\
&\langle w_1,m_1\rangle
=\langle w_2,m_2\rangle=2,\nonumber\\
&\langle w_1,m_2\rangle=-1,\quad
\langle w_2,m_1\rangle=-2.
\end{align}
$w_i$ and $m_i$ are regarded as simple roots of $so(5)$ and
the simple roots of the dual algebra $usp(4)={}^Lso(5)$.
$w_2$ and $m_1$ are long, and $w_1$ and $m_2$ are short.
(Figure \ref{so5dyon.eps})
The vectors span the lattice $\Lambda_{\rm dyn}$ of dynamical particles,
and the charges of external line operators up to screening are
classified by $\Lambda/\Lambda_{\rm dyn}=\ZZ_2$.

Without taking account of the screening effect,
the charge carried by an external fundamental string is (Figure \ref{so5dyon.eps})
\begin{align}
q_{\rm F1}=(1,0;0,0)=w_1+\frac{1}{2}w_2+\frac{1}{2}m_1+m_2.
\end{align}
Again, this charge cannot be transformed into that of a pure Wilson line operator
by a duality transformation without using screening.
To calculate the index, we substitute the following $\chi_m$ and $\chi_e$ into
the formula (\ref{ibr}).
\begin{align}
\chi_m
&=q^{1/2}(z_1^2+z_1z_2+z_2^2)
+\tfrac{z_1}{z_2}+2+\tfrac{z_2}{z_1}
+q^{1/2}(\tfrac{1}{z_1^2}+\tfrac{1}{z_1z_2}+\tfrac{1}{z_2^2})),\nonumber\\
\chi_e
&=z_1+z_2.
\label{so5chmche}
\end{align}
The result is
\begin{align}
I_{SO(5),q_{\rm F1}}
&=1+(2u+\tfrac{2}{u})t+(3u^2+3+\tfrac{3}{u^2})t^2+(4u^3+2u+\tfrac{2}{u}+\tfrac{4}{u^3})t^3+\cdots
\end{align}
We observe an improvement
after including the giant graviton contributions.
Although the agreement is less precise than
in the $k=3$ case,
it remains reasonably good.

\subsubsection{$k=6$}
The S-fold theory $S_{6,1}(2)$
is expected to be equivalent to ${\cal N}=4$ SYM
with the $G_2$ gauge group.
This case is qualitatively different
because the candidate gauge-theory description
involves non-minuscule magnetic charges,
for which monopole bubbling contributions are essential.
Therefore the comparison performed in this subsection should be
regarded as preliminary.

\begin{figure}[htb]
\centering
\includegraphics{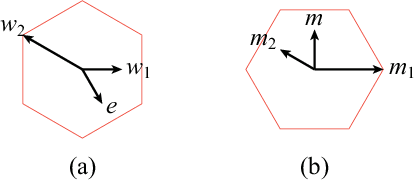}
\caption{The electric component $e$ and the magnetic component $m$ of the charge $q_{\rm F1}$ are
shown as vectors in the weight lattices of $G_2$ and $G_2={}^LG_2$.}\label{g2dyon.eps}
\end{figure}

For $k=6$, we can take the following basis vectors of dynamical charges
\begin{align}
w_1&=(1,0;-1,0),&
w_2&=(-1,-1;2,-1),\nonumber\\
m_1&=(1,1;-1,-1),&
m_2&=(0,-1;1,0).
\end{align}
The Dirac pairings are
\begin{align}
\langle w_1,w_2\rangle=
\langle m_1,m_2\rangle=0,\nonumber\\
\langle w_1,m_1\rangle=
\langle w_2,m_2\rangle=2,\nonumber\\
\langle w_1,m_2\rangle=-1,\quad
\langle w_2,m_1\rangle=-3.
\end{align}
$w_i$ and $m_i$ are regarded as simple roots of
the electric $G_2$ and the magnetic $G_2={}^LG_2$.
$w_1$ and $m_2$ are short, and $w_2$ and $m_1$ are long.
The fundamental string charge is expanded as (Figure \ref{g2dyon.eps})
\begin{align}
q_{\rm F1}=(1,0;0,0)=-w_1-w_2+m_1+2m_2
\label{qf1g2}
\end{align}
Again, the charges are dyonic.
In particular, the monopole charges are non-vanishing.

We first give the result
for ${\cal N}=4$ SYM with $G=G_2$ without a line-operator insertion.
\begin{align}
I_{G_2}
&=1+0t+(u^2+1+\tfrac{1}{u^2})t^2
+(-2u-\tfrac{2}{u})t^3
\nonumber\\&
+(u^4+2u^2+4+\tfrac{2}{u^2}+\tfrac{1}{u^4})t^4+\cdots.
\end{align}
The corresponding results on the gravity side are \cite{Arai:2019xmp}
\begin{align}
I_{S_{6,1}(2)}^{(m=0)}
&=1+0t+t^2+(-2u-\tfrac{2}{u})t^3
+4t^4+\cdots,\nonumber\\
I_{S_{6,1}(2)}^{(m\leq1)}
&=1+0t+(u^2+1+\tfrac{1}{u^2})t^2+(-2u-\tfrac{2}{u})t^3
\nonumber\\&
+(2u+4+\tfrac{2}{u})t^4+\cdots
\end{align}
The agreement improves once the giant graviton
contributions are included.

Let us turn to the line operator indices.
The formula (\ref{thooftindex}) and (\ref{ibr}) are applicable
only when the magnetic charge is the highest weight of a minuscule representation.
The only minuscule representation of $G_2$
is the trivial representation, which is not the case here.
To calculate the index, we need to take into account the monopole bubbling effect.
For example,
let us see the index for a 't Hooft line with
magnetic charge $m=m_2$.
The formula (\ref{thooftindex}) gives
\begin{align}
I_{G_2,{\rm mon}}^{(0)}
&=1+(u+\tfrac{1}{u})t+(2u^2+1+\tfrac{2}{u^2})t^2
+(2u^3+\tfrac{2}{u^3})t^3
\nonumber\\&
+(3u^4+u^2+1+\tfrac{1}{u^2}+\tfrac{3}{u^4})t^4+\cdots
\label{g20}
\end{align}
The superscript `$(0)$' indicates that this does not take account of the
monopole bubbling effects.
Due to the S-duality,
the index for the 't Hooft line agrees with the result
for the fundamental Wilson line index with $e=w_1$
\begin{align}
I_{G_2,{\rm fund}}
&=1+(u+\tfrac{1}{u})t+(2u^2+1+\tfrac{2}{u^2})t^2+(2u^3+\tfrac{2}{u^3})t^3
\nonumber\\&
+(3u^4+u^2+2+\tfrac{1}{u^2}+\tfrac{3}{u^4})t^4+\cdots.
\end{align}
The discrepancy starting at order $t^4$
is interpreted as the result of the omission of the
monopole bubbling effects in (\ref{g20}).
This comparison
suggests that the first few terms
remain reliable even without including monopole bubbling.

For the Wilson-'t Hooft line with charge (\ref{qf1g2}),
the result without monopole bubbling is
\begin{align}
I_{G_2,q_{\rm F1}}^{(0)}
&=1+(2u+\tfrac{2}{u})t+(3u^2+3+\tfrac{3}{u^2})t^2
\nonumber\\&
+(4u^3+2u+\tfrac{2}{u}+\tfrac{4}{u^3})t^3
+(5u^4+2u^2-1+\tfrac{2}{u^2}+\tfrac{5}{u^4})t^4+\cdots
\end{align}
The corresponding AdS-side result
without giant graviton contributions is
\begin{align}
I_{S_{6,1},{\rm F1}}^{(m=0)}
&=1+(u+\tfrac{1}{u})t+(u^2+1+\tfrac{1}{u^2})t^2+(u^3+\tfrac{1}{u^3})t^3
\nonumber\\&
+(u^4+2+\tfrac{1}{u^4})t^4+\cdots.
\end{align}
Including the giant graviton contributions,
we obtain
\begin{align}
I_{S_{6,1},{\rm F1}}^{(m\leq1)}
&=1+(2u+\tfrac{2}{u})t+(2u^2+3+\tfrac{2}{u^2})t^2
+(2u^3+u+\tfrac{1}{u}+\tfrac{2}{u^3})t^3
\nonumber\\&
+(2u^4+u^2+\tfrac{1}{u^2}+\tfrac{2}{u^4})t^4+\cdots.
\end{align}
The discrepancy first appears at $O(t^2)$.
It would be interesting
to determine whether
this agreement can be improved
by including monopole
bubbling contributions on the gauge theory side
and
multiple-wrapping contributions
on the AdS side.

\section{Fat strings}\label{fat.sec}
In this section we study fat-string type line operators
realized by fivebranes wrapped around four-cycles
in the internal space.
Our main goal is to determine their contribution
to the Schur index for all $k$.
Furthermore, for $k=2$, we compare the results with
the results on the gauge theory side.

For $U(N)$ SYM ($=S_{1,1}(N)$),
a fivebrane wrapped around an $S^4\subset S^5$
corresponds to line operators in the totally anti-symmetric representations \cite{Yamaguchi:2006tq}.
The angular radius $\theta$ of the $S^4$
is determined by the rank $n$ of the representation
(the number of boxes in the Young diagram)
by the relation
\begin{align}
\frac{n}{N}=\frac{1}{\pi}(\theta-\sin\theta\cos\theta).
\end{align}
Mode analysis of the fields on the fivebrane was performed
in \cite{Faraggi:2011bb,Faraggi:2011ge}.
(See Appendix \ref{modesond5.app}.)
It was confirmed in \cite{Gang:2012yr}
that the line-operator index in the large $N$ and large $n$ limit
is correctly reproduced as the product of the
supergravity index and the index of fluctuation modes on the fivebrane,
The finite-$N$ and finite-$n$ index is given by the
giant graviton expansion \cite{Imamura:2024pgp}.
\begin{align}
I=I_{\rm sugra}I_{\rm D5}\sum_{m,m'}x^{mn+m'(N-n)}F_{m,m'}
\end{align}
where we assume the $y$-series expansion
and include only the contribution from giant gravitons on $S_X$.
The cycle $S_X$ is divided into two parts $S_X^{(1)}$ and $S_X^{(2)}$
at the intersection with the fivebrane,
and $m$ and $m'$ are the numbers of the D3-branes
wrapped on these two parts.
The factor $x^{mn+m'(N-n)}$ reflects the energy of
wrapped D3-branes and
plays the role of the counting parameter
in the giant graviton expansion.
(Strictly speaking, ``giant gravitons'' originally refer to
objects
that carry the same quantum numbers as pointlike gravitons,
but we will abuse the term for disk D-branes.)

\subsection{$k=2$ case}
In the case of $k=2$, $\ZZ_2$ action flips the
orientation in the $\theta$ direction, and $\theta$ is
transformed to $\pi-\theta$.
A fat string we will discuss in this section is
a D5-brane at the equator $\theta=\pi/2$,
which is identified by the $\ZZ_2$ projection
with itself \cite{Witten:1998xy}.
Each point on the D5-brane is identified with its antipodal point.
Namely, the D5-brane is wrapped on the non-trivial four cycle
with the topology $RP^4$.

For giant gravitons, because two disks $S_X^{(1)}$ and $S_X^{(2)}$ are identified,
the expansion reduces to the simple expansion
\begin{align}
I_{S_{2,1}(N),{\rm fivebrane}}=I_{\rm sugra}I_{\rm fivebrane}\sum_{m}x^{mN}F_m
\label{ggfivebrane}
\end{align}
where $I_{\rm fivebrane}$ is the contribution from the fluctuation modes
on the fivebrane.
It is easily obtained by the $\ZZ_2$ projection
as
\begin{align}
I_{\rm fivebrane}=\pexp({\cal P}_2f_{\rm fivebrane}),
\end{align}
where $f_{\rm fivebrane}$ is the refined index of the
vector multiplet on the fivebrane.
\begin{align}
f_{\rm fivebrane}=\eta\frac{1-\eta q}{(1-x)(1-y)}.
\label{fd5refined}
\end{align}
See Appendix \ref{modesond5.app} for derivation.

The description of the theory on giant gravitons
depends on the duality frame,
and any description should give the same giant graviton indices $F_m$.

\paragraph{D3-NS5 system}
Let us first discuss the duality frame
in which the fivebrane is an NS5-brane.
The D3-brane worldvolume $S^3$ is divided into
a pair of hemispheres, which are identified by
the orientifold.
The Neumann type boundary condition
is imposed on the fields on each hemisphere \cite{Gaiotto:2008sa,Gaiotto:2008ak}.
There exist three-dimensional hypermultiplets
on the equator arising from open strings
attached to D3-branes on two sides of the
equator.
The gauge group is $U(m)$,
and the hypermultiplets belong to
the symmetric and anti-symmetric representations
of $U(m)$.

The index of a vector multiplet
on the hemisphere is called a half index
\cite{Dimofte:2011ju,Gadde:2013wq,Okazaki:2019ony}.
The letter index is
\begin{align}
f_{\rm half}=\frac{x-q}{1-q}.
\label{fhalf}
\end{align}
The refined letter index of the three-dimensional hypermultiplet
on $S^3$ is
\begin{align}
f_{\rm hyper}=\frac{y^{\frac{1}{2}}(1-\eta x)}{1-q}.
\label{fhyper}
\end{align}
(See \cite{Kim:2009wb,Imamura:2011su} for unrefined letter indices of 3d multiplets.)
(\ref{fhalf}) and 
(\ref{fhyper}) are given as the indices of
boundary theories,
and to obtain the giant graviton contribution,
we need to apply the variable changes (\ref{sxsy})
before the projection.
Namely,
$F_m$ is given by
\begin{align}
F_m=
\oint_{U(m)} dU \pexp( & f_{\rm half} {\chi}_{\text{adj}}^{U(m)}(U) \nonumber \\ 
&+ {\cal P}_2\sigma_x (
    f_{\rm hyper}(
      {\chi}_{\text{sym}}^{U(m)}(U)+ \eta {\chi}_{\text{antisym}}^{U(m)}(U) \nonumber \\
      &\hspace{15mm}+ {\chi}_{\text{sym}}^{U(m)}(U^{-1})+ \eta {\chi}_{\text{antisym}}^{U(m)}(U^{-1})
    )
  )).
\label{f10a}
\end{align}
The letter index ${\cal P}_2\sigma_xf_{\rm hyper}$ is also obtained
by applying $\sigma_x$ to
the index of a hypermultiplet in $\mathbb{R} \times RP^2$ derived in \cite{Tanaka:2014oda}.

\paragraph{D3-D5 system}

In the frame with a D5-brane,
the theory on $m$ D3-branes
is the theory on $RP^3$
coupled to a twisted hypermultiplet
on the equator.
The twisted hypermultiplet arises from
open strings between D3-branes and the D5-brane.
The gauge group on $RP^3$ is locally $U(m)$,
and is broken by the $\ZZ_2$ identification to $O(m)$.
The twisted hypermultiplet belongs to
the vector representation.

The letter index of a twisted hypermultiplet is
\begin{align}
f_{\rm twisted}=\frac{x^{\frac{1}{2}}(1-y)}{1-q}.
\end{align}
The giant graviton index $F_m$ is given by
\begin{align}
F_m= \int_{O(m)}dU \pexp({\cal P}_2\sigma_x f_{\rm vec} {\chi}_{\text{adj}}^{U(m)}
  +\sigma_xf_{\rm twisted} {\chi}_{\rm vec}^{O(m)}).
\label{f10b}
\end{align}
The D3-NS5 system and the D3-D5 system are related by the S-duality,
and the two integrals
(\ref{f10a}) and (\ref{f10b}) give the same giant graviton indices.

\begin{figure}[htb]
\centering
\includegraphics[width=1\textwidth]{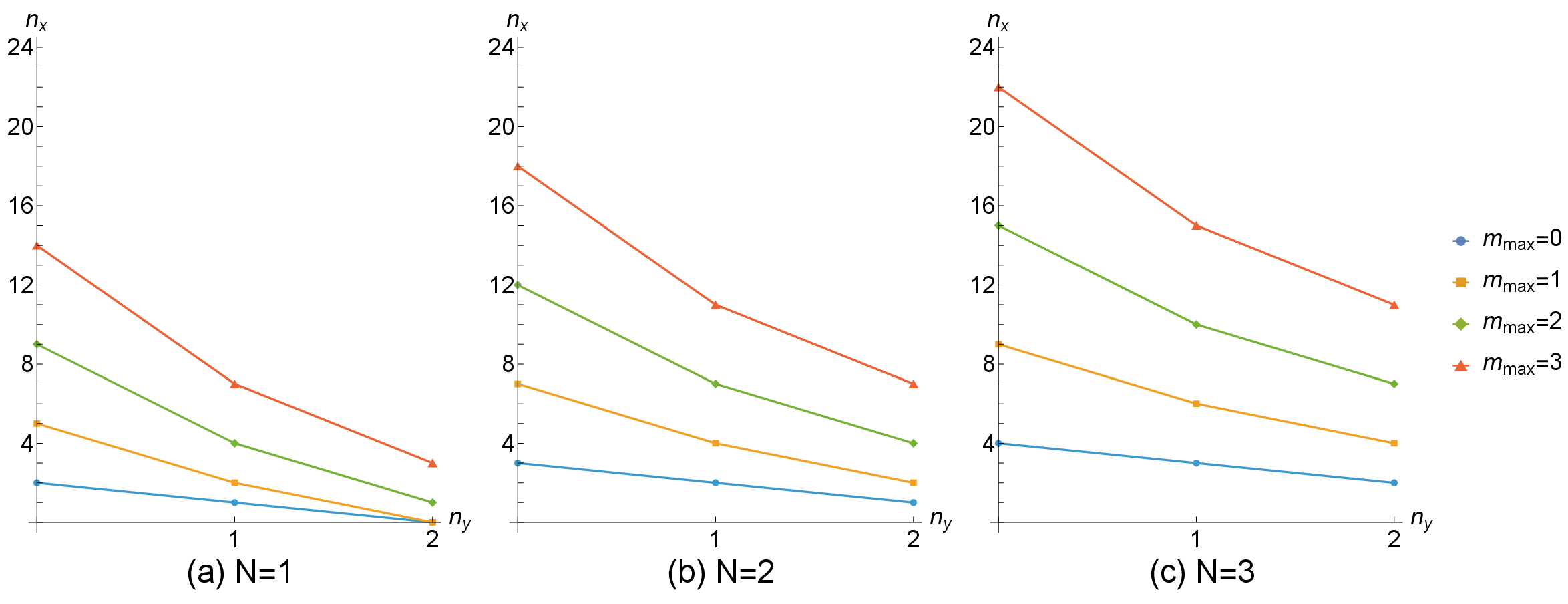}
\caption{The leading order in $x$ at each order of $y$ in the expansion of ${\Delta}_{SO(2N),{\rm spinor}}^{(m_{\text{max}})}$ up to $m_{\text{max}} \le 3$: 
$n_x$ is the leading order in $x$, and $n_y$ is the order in $y$.}\label{DeltaSO2Nspi}
\end{figure}

\paragraph{Numerical checks}
Let us numerically check that the expansion
(\ref{ggfivebrane}) works.
We calculate the index using (\ref{ggfivebrane}) with the cutoff $m_{\rm max}$.
Namely, we include only contributions with the wrapping number $m\leq m_{\rm max}$
and denote the result by
$I_{S_{2,1}(N),{\rm fivebrane}}^{(m\leq m_{\rm max})}$.
We compare the results with the indices with the spinor line insertion calculated on the
gauge theory side.
We expand the discrepancies into $y$-series.
\begin{align}
  {\Delta}_{SO(2N),{\rm spinor}}^{(m)} = {\cal I}_{SO(2N),{\rm spinor}} - {\cal I}_{S_{2,1}(N),{\rm fivebrane}}^{(m\leq m_{\rm max})}
=\sum_{n=0}^\infty c_n(x)y^n.
\end{align}
If the giant graviton expansion works,
the orders of $c_n(x)$ should increase as we increase the cut-off $m_{\rm max}$.
The results for small ranks $N=1,2,3$ are shown in Figure \ref{DeltaSO2Nspi}.
We find the expected behavior of the orders of $c_n(x)$.

\subsection{Fat strings for $k\geq 3$}

Let us consider
fat-string type line operators in the S-fold theories with $k=3,4,6$.
Since the boundary-theory description is not understood,
we focus on the gravity-side analysis.
Furthermore, there is a difficulty in
calculating the contribution from
giant gravitons, as we will mention later,
and we consider
only the contribution from fivebrane configurations,
which is expected to give the line operator
index in the large $N$ limit.

\begin{figure}[htb]
\centering
\includegraphics{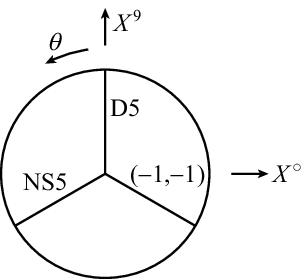}
\caption{A fat string of the $\ZZ_3$ S-fold theory}\label{fatz3.eps}
\end{figure}

First, we need to identify the fivebrane BPS configurations in
$S^5/\ZZ_k$
associated with the non-trivial elements of
$\Gamma_k^{(5)}$ in (\ref{gk1gk5}).
The four cycle associated with the generating element
of $\Gamma_k$ is
constructed by combining
a fivebrane wrapped on a four-dimensional
hemisphere in $S^5$ and its $\ZZ_k$ images.
We show the configuration for $k=3$ in Figure \ref{fatz3.eps}.
It shows the projection of the configuration
to $Z$-plane.
The $k$ radial segments represent
fivebranes wrapped over hemispheres,
and they meet at the junction locus
projected to the origin.
They are related by the S-fold
$\ZZ_k$ action, and their $(p,q)$ charges
satisfy the conservation law
at the junction.
These configurations preserve the same supersymmetries as the
ones for the fundamental lines, and are $1/3$ BPS.
See Appendix \ref{killing.app}.

One can confirm that the configuration carries $\Gamma_k^{(5)}$ valued
charges.

For $k=3$, $\Gamma_3^{(5)}=\ZZ_3$, and three copies of fat strings
can be annihilated by continuous deformation.
This is shown as follows.
Let us start with three copies of
coincident fat string configurations shown in Figure \ref{fatz3.eps}.
We rotate each of them by angles $0$, $\frac{2\pi}{3}$, and $\frac{4\pi}{3}$.
After the rotation,
the charges of the three branes coincident along each branch sum up to zero,
and they can annihilate.
This implies that a copy of the fat string configuration carries a $\ZZ_3$-valued charge.

For $k=4$, $\Gamma_4^{(5)}=\ZZ_2$, and we start from two copies
of coincident fat string configurations.
We rotate one of them by an angle $\pi$.
The two fat strings then carry opposite fivebrane charges
on each branch and can annihilate in pairs.
This implies that a copy of a fat string carries a $\ZZ_2$-valued charge.

For $k=6$, $\Gamma_6=0$, and
a single fat string has vanishing topological charge.
This can be shown as follows.
We start from a single fat string configuration.
We can decompose each branch with complex charge $q_C=p+\tau q$
($\tau=e^{2\pi i/6}$)
into two fivebranes with charge $\tau q_C$ and $\tau^{-1}q_C$
because $\tau+\tau^{-1}=1$.
As a result, one fat string is decomposed into two fat strings.
We can rotate one of them
by $\pi/6$ so that the charges of the coincident branes on each oranch,
carry opposite charges,
and they can annihilate in pairs.
This implies that the original fat string configuration does not carry a conserved charge.
Although a $\ZZ_6$ fat string
does not carry a non-trivial topological charge,
it is BPS and contributes to
the line operator index.

The contribution of the fat-string configurations to the line operator index
can be obtained by
analyzing the fluctuation modes
on the junction
with appropriate
boundary conditions imposed on the junction locus,
and summing up their contributions to the
index.
We describe the details of the
calculation in Appendix \ref{fat.app},
and here we show only the final result.
The letter index of the fluctuation modes on the
fat-string configuration is
\begin{align}
i_k=
\frac{1}{1-q}\left(\frac{x}{1-x}+\frac{y}{1-y}+1\right)
-\frac{1+q}{1-q}\left(\frac{x^k}{1-x^k}+\frac{y^k}{1-y^k}+1\right).
\label{fatkletter0}
\end{align}
The line operator index for the fat string in the large $N$ limit is given by
\begin{align}
I_k=\pexp({\cal P}_kf_{\rm sugra})\pexp i_k.
\end{align}
The first few terms in the $t$-expansion for each $k$ are as follows:
\begin{align}
I_2&=1+2t+5t^2+8t^3+20t^4+26t^5+62t^6+70t^7+176t^8+\cdots,\nonumber\\
I_3&=1+2t+5t^2+10t^3+22t^4+40t^5+77t^6+136t^7+244t^8+\cdots,\nonumber\\
I_4&=1+2t+5t^2+10t^3+24t^4+42t^5+89t^6+150t^7+299t^8+\cdots,\nonumber\\
I_6&=1+2t+5t^2+10t^3+24t^4+44t^5+93t^6+164t^7+323t^8+\cdots.
\label{ifatk}
\end{align}

\section{Summary and Discussion}\label{summary.sec}

In this paper, we studied superconformal indices with line operator insertions
in S-fold theories.
On the gravity side, line operators are realized by fundamental strings and
wrapped fivebranes in AdS$_5 \times S^5/\mathbb{Z}_k$.

For the $\mathbb{Z}_2$ S-fold, corresponding to orientifold theories,
we compared the holographic results with line operator indices
of the vector representation and the spinor representation
in $\mathcal{N}=4$ super Yang--Mills theories with orthogonal gauge groups.
We found that the inclusion of giant graviton contributions
improves the agreement between the gravity-side and gauge-theory results.

We also studied the intrinsically strongly coupled cases with
$k=3,4,6$.
Although direct calculations in the boundary theories are difficult,
the holographic analysis yields concrete predictions for line operator
indices associated with fundamental strings and wrapped fivebranes.
For the fundamental line operators in rank-two theories,
the resulting indices are consistent with the proposed enhancement
from ${\cal N}=3$ to ${\cal N}=4$ supersymmetry.
For the fat-string type line operators
we constructed BPS brane configurations
using fivebrane junctions and calculated the
index
by the mode analysis on the fivebrane junctions.
These results are predictions for the
corresponding line-operator index in the large $N$ limit.

There remain several important open problems.
One issue is the contribution from multiply wrapped giant gravitons.
In the present work, we mainly considered singly wrapped configurations.
A systematic treatment of multiply wrapped giant graviton
configurations would provide a more complete
understanding of the finite-$N$ expansion.

Another important issue is monopole bubbling.
For several line operators, especially in the $k=6$ case, the corresponding
charges are dyonic, and monopole bubbling contributions are
expected to play a crucial role.
It would be interesting to develop methods for incorporating these effects
and to compare the resulting indices with the holographic predictions.

It would also be interesting to better understand the boundary-theory
interpretation of wrapped fivebrane configurations in S-fold theories with
$k\geq 3$.
Because these theories do not admit known Lagrangian descriptions,
line operators may provide a useful probe of their dynamics.
The fat-string indices studied in this paper provide
dynamical information beyond the topological classification
of line operators.

We hope that the present analysis demonstrates that
line-operator indices provide a useful new probe of
S-fold theories, complementary to the classification
based on generalized global symmetries, and contributes
to a better understanding of defect operators and
finite-$N$ effects in holography for non-Lagrangian
superconformal field theories.

\section*{Acknowledgments}
The authors thank T.~Nishinaka and D.~Yokoyama for valuable comments.
The work of Y.~I. was supported by JSPS KAKENHI Grant Number JP26K07103,
the work of M.~I. was supported by JST SPRING, Grant Number JPMJSP2180,
and the work of A.~S. was supported by JSPS KAKENHI Grant Number JP24KJ1105.

\appendix
\section{Killing spinors and supercharges}\label{killing.app}
Let us define ambient spaces $\RR^{2,4}$ and $\RR^6$ in which
$AdS_5$ and $S^5$ are embedded, respectively.
To make the $so(2,4)\times so(6)$ symmetry manifest,
we introduce orthonormal coordinates $x^M$ ($M=\bullet01234$) for $\RR^{2,4}$ and
$x^A$ ($A=\circ56789$) for $\RR^6$.
Let $T^{MN}$ and $T^{AB}$ be the generators of $so(2,4)$ and $so(6)$,
respectively.
We define the Cartan generators as follows.
\begin{align}
&H=T^{\bullet0},\quad
J_1=T^{12},\quad
J_2=-T^{34},\nonumber\\
&R_1=T^{56},\quad
R_2=T^{78},\quad
R_3=T^{\circ9}.
\end{align}

We use the mostly plus convention for the metric.
Let $\gamma^M$ ($M=0\cdots9$) be ten-dimensional Dirac matrices,
and $\ol\gamma=\gamma^{0123456789}$ be the chirality matrix.
Let $\varepsilon=(\varepsilon_1,\varepsilon_2)$
be the supersymmetry transformation parameter
of type IIB supergravity.
Each $\varepsilon_I$ ($I=1,2$) is a Majorana-Weyl spinor with
positive chirality; $\ol\gamma\varepsilon_I=\varepsilon_I$.
Let $SO(2)_A=U(1)_A$ be the R-symmetry
rotating the two Majorana-Weyl spinors $\varepsilon_I$,
which is broken by the flux quantization.
Let $\sigma_i$ ($i=x,y,z$) be the Pauli matrices
acting on the $SO(2)_A$ indices $I,J,\ldots$.
The gravitinos in the type IIB supergravity multiplet have positive chirality,
and their local supersymmetry transformation law is\footnote{We use the notation $\sla A_n=\frac{1}{n!}\gamma^{M_1\cdots M_n}A_{M_1\cdots M_n}$ for
an $n$-form field $A_n$.}
\begin{align}
\delta\psi_M=D_M\varepsilon+\frac{g_{\rm str}}{16}i\sigma_y(\sla G_5\gamma_M)_L\varepsilon.
\label{deltapsi}
\end{align}
In (\ref{deltapsi}) the three-form fields are set to be zero,
and the scalar fields are assumed to be constants.
$G_5$ is the five-form flux, and
$(\cdots)_L$ represents the positive-chirality part of the matrix.
In the $AdS_5\times S^5$ near horizon geometry
the second term of (\ref{deltapsi}) is given by
\begin{align}
\frac{g_{\rm str}}{16}(i\sigma_y\sla G)_L=-\frac{1}{2L}(\gamma^\bullet\gamma_M)_L,\quad
\gamma^\bullet=\pm i\sigma_y\gamma^{56789}.
\label{fluxg5}
\end{align}
where the sign in the definition of $\gamma^\bullet$ depends on the
orientation of the D3-branes, and we will fix the ambiguity
in a way consistent with our definition of the superconformal index.
Namely, we choose a sign convention with which
the supercharge with the quantum numbers
\begin{align}
(H,J_1,J_2,R_1,R_2,R_3,A)=(
+\tfrac{1}{2},
-\tfrac{1}{2},
-\tfrac{1}{2},
+\tfrac{1}{2},
+\tfrac{1}{2},
+\tfrac{1}{2},
-\tfrac{1}{2})
\label{qinsci}
\end{align}
exists.
As we will show shortly, we need to choose
the negative sign in (\ref{fluxg5}).

A Killing spinor is a solution to the Killing spinor equation $\delta\psi_M=0$.
Substituting (\ref{fluxg5}) into the Killing spinor equation, we obtain
\begin{align}
AdS_5&: 0=\delta\psi_\mu=D_\mu\varepsilon-\frac{1}{2L}\gamma^\bullet\gamma_\mu\varepsilon \quad(\mu=01234),\nonumber\\
S^5&: 0=\delta\psi_m=D_m\varepsilon-\frac{1}{2L}\gamma^\bullet\gamma_m\varepsilon\quad(m=56789).
\label{killingeqs}
\end{align}
Because the Killing spinor equations are first-order differential equations,
a solution is specified by the value of $\varepsilon$ at an arbitrary point.
Let us take $P_0$ with the coordinates
\begin{align}
x^\bullet=x^\circ=L,\quad
\mbox{others}=0
\end{align}
as a reference point.
We can identify $\varepsilon$ at $P_0$ as the parameter $\xi$ appearing in the
supersymmetry transformation
\begin{align}
\delta{\cal O}=[{\cal O},i\xi Q].
\end{align}
In a neighborhood of $P_0$ we can use $x^\mu$ ($\mu=01234$) and $x^m$ ($m=56789$)
as orthonormal coordinates.
Under the stabilizer subgroup of $P_0$, 
$\xi=\varepsilon|_{P_0}$ is transformed by the representation matrices
\begin{align}
\rho(T^{\mu\nu})=-\tfrac{i}{2}(\gamma^{\mu\nu})_L,\quad
\rho(T^{mn})=-\tfrac{i}{2}(\gamma^{mn})_L.
\label{repxi1}
\end{align}
To describe the action of the full isometry group,
we need to extend (\ref{repxi1}) to
$T^{\bullet\mu}$ and $T^{\circ m}$
so that they satisfy
the $so(2,4)\times so(6)$ algebra.
This can be achieved by using
the matrices appearing in the second terms of equations in 
(\ref{killingeqs}).
Namely, we define
\begin{align}
\rho(T^{\bullet\mu})
=-\rho(T^{\mu\bullet})
=-\tfrac{i}{2}(\gamma^\bullet\gamma^\mu)_L,\quad
\rho(T^{\circ m})
=-\rho(T^{m\circ})
=-\tfrac{i}{2}(\gamma^\bullet\gamma^m)_L.
\label{repxi2}
\end{align}
It is easy to confirm that $\rho(T^{MN})$ and $\rho(T^{AB})$ satisfy
the commutation relations of the $so(2,4)\times so(6)$ algebra.
With these matrices we can rewrite the Killing spinor equations in
(\ref{killingeqs}) as
\begin{align}
AdS_5&: 0=D_\mu\varepsilon-\frac{i}{L}\rho(T^\bullet{}_\mu)\varepsilon=D_\mu^{\RR^{2,4}}\varepsilon,\nonumber\\
S^5&: 0=D_m\varepsilon-\frac{i}{L}\rho(T^\circ{}_m)\varepsilon=D_m^{\RR^6}\varepsilon,
\label{killingeqs2}
\end{align}
where $D_\mu^{\RR^{2,4}}$ and $D_m^{\RR^6}$ are
covariant derivatives in the ambient spaces.
(\ref{killingeqs2}) implies that a Killing spinor $\varepsilon$ can be regarded as a
constant bi-spinor in the ambient space $\RR^{2,4}\times \RR^6$.
Hence we can identify $\varepsilon$ with $\xi$.
The parameter corresponding to the supercharge with the quantum numbers
(\ref{qinsci}) satisfies
\begin{align}
&\rho(H)\xi=
-\rho(J_1)\xi=
-\rho(J_2)\xi=
\nonumber\\&
\rho(R_1)\xi=
\rho(R_2)\xi=
\rho(R_3)\xi=
-\rho(A)\xi=
\tfrac{1}{2}\xi.
\end{align}

The supercharge $Q$ belongs to the dual representation of the representation of $\xi$,
and is transformed so that the product $(\xi Q)$ is invariant.
The representation matrices $\rho'$ for $Q$ are
given by
\begin{align}
\rho'=-\gamma^\bullet\rho\gamma^\bullet.
\end{align}
In particular,
the representation matrices for the Cartan generators are
\begin{align}
&\rho'(H)=+\tfrac{i}{2}(\gamma^\bullet\gamma^0)_R,\quad
\rho'(J_1)=-\tfrac{i}{2}(\gamma^{12})_R,\quad
\rho'(J_2)=+\tfrac{i}{2}(\gamma^{34})_R,\quad
\nonumber\\
&\rho'(R_1)=-\tfrac{i}{2}(\gamma^{56})_R,\quad
\rho'(R_2)=-\tfrac{i}{2}(\gamma^{78})_R,\quad
\rho'(R_3)=-\tfrac{i}{2}(\gamma^\bullet\gamma^9)_R,
\nonumber\\
&\rho'(A)=-\tfrac{1}{2}\sigma_y.
\end{align}
The following relations hold.
\begin{align}
16
\rho'(H)
\rho'(J_1)
\rho'(J_2)
\rho'(A)
=
16
\rho'(R_1)
\rho'(R_2)
\rho'(R_3)
\rho'(A)
=\pm1.
\end{align}
The sign on the right hand side is the same as the
sign in (\ref{fluxg5}).
For the existence of the supercharge with the quantum numbers in
(\ref{qinsci}) this must be $-1$,
and we need to choose the negative sign in (\ref{fluxg5}).

Let $Q^I_a$ and $\ol Q_I^{\dot a}$ be the supercharges
($I=1,2,3,4$, $a=\pm$, $\dot a=\dot\pm$).
We use the convention with the charge assignments of
these Cartan generators and the $U(1)_A$ charge $A$ shown in Table \ref{qcharges.tbl}.
\begin{table}[htb]
\caption{The quantum numbers of the supercharges}\label{qcharges.tbl}
\centering
\begin{tabular}{c|c|cc|ccc|c}
                & $H$            & $J_1$            & $J_2$            & $R_x$          & $R_y$          & $R_z$          & $A$ \\
\hline
$Q_\pm^{1}$ & $+\frac{1}{2}$ & $\pm\frac{1}{2}$ & $\mp\frac{1}{2}$ & $-\frac{1}{2}$ & $-\frac{1}{2}$ & $-\frac{1}{2}$ & $+\frac{1}{2}$ \\
$Q_\pm^{2}$ &                &                  &                  & $-\frac{1}{2}$ & $+\frac{1}{2}$ & $+\frac{1}{2}$ &                \\
$Q_\pm^{3}$ &                &                  &                  & $+\frac{1}{2}$ & $-\frac{1}{2}$ & $+\frac{1}{2}$ &                \\
$Q_\pm^{4}$ &                &                  &                  & $+\frac{1}{2}$ & $+\frac{1}{2}$ & $-\frac{1}{2}$ &                \\
\hline
$\ol Q^{\dot\pm}_{1}$
                & $+\frac{1}{2}$ & $\mp\frac{1}{2}$ & $\mp\frac{1}{2}$ & $+\frac{1}{2}$ & $+\frac{1}{2}$ & $+\frac{1}{2}$ & $-\frac{1}{2}$ \\
$\ol Q^{\dot\pm}_{2}$ &      &                  &                  & $+\frac{1}{2}$ & $-\frac{1}{2}$ & $-\frac{1}{2}$ &  \\
$\ol Q^{\dot\pm}_{3}$ &      &                  &                  & $-\frac{1}{2}$ & $+\frac{1}{2}$ & $-\frac{1}{2}$ &  \\
$\ol Q^{\dot\pm}_{4}$ &      &                  &                  & $-\frac{1}{2}$ & $-\frac{1}{2}$ & $+\frac{1}{2}$ &  \\
\end{tabular}
\end{table}

\paragraph{Line insertion}
The insertion of line operators breaks a part of the supersymmetry.
We consider different $1/2$ BPS line operators, which preserves the same half of the supercharges.
Let us consider a line operator realized by a fundamental string,
and assume that its worldsheet contains the reference point $P_0$ and extends along $04$ directions at $P_0$.
The projection condition imposed on the parameter $\xi=\varepsilon|_{P_0}$ for the preserved rigid supersymmetry is given by
\begin{align}
\gamma^{04}\sigma_z\xi=+\xi.
\label{f04}
\end{align}

Because $\gamma^{04}\sigma_z$ anti-commutes with $\rho(J_2)$, $\rho(R_3)$, and $\rho(A)$,
and commutes with $\rho(H)$, $\rho(J_1)$, $\rho(R_1)$, and $\rho(R_2)$,
(\ref{f04}) relates a component of $\xi$ to another component with
opposite $J_2$, $R_3$, and $A$ quantum numbers.
Correspondingly, only the following linear combinations
of supercharges are preserved.
\begin{align}
q^I_\pm=Q^I_\pm-\ol Q^{\dot\mp}_{5-I}.
\end{align}
See Table \ref{qqn.tbl} for quantum numbers of $q^I_a$.
\begin{table}[htb]
\caption{Unbroken supercharges in the line-operator background}\label{qqn.tbl}
\centering
\begin{tabular}{c|c|c|cc}
                & $H$            & $J_1$            & $R_x$          & $R_y$        \\
\hline
$q_\pm^{\hat1}$ & $+\frac{1}{2}$ & $\pm\frac{1}{2}$ & $-\frac{1}{2}$ & $-\frac{1}{2}$ \\
$q_\pm^{\hat2}$ &                &                  & $-\frac{1}{2}$ & $+\frac{1}{2}$ \\
$q_\pm^{\hat3}$ &                &                  & $+\frac{1}{2}$ & $-\frac{1}{2}$ \\
$q_\pm^{\hat4}$ &                &                  & $+\frac{1}{2}$ & $+\frac{1}{2}$ \\
\end{tabular}
\end{table}

\paragraph{S-fold projection}

For $k\geq 3$, the $\ZZ_k$ projection with the generator
(\ref{zkgenerator})
eliminates $Q_a^2$ and $\ol Q^{\dot a}_2$,
and as the result, we obtain a system with
a genuine ${\cal N}=3$ supersymmetry.
In the presence of a line operator,
$q_a^{\hat 2}$ and $q_a^{\hat 3}$ are broken by the S-folding,
but still $q_a^{\hat 1}$ and $q_a^{\hat 4}$ are preserved.
Out of $12$ supercharges of $S$-fold theory
the line operator preserves $4$ of them,
and the line is $1/3$ BPS.

\section{D5 brane on $S^4$}\label{d5action.app}
\subsection{Action}
Let us consider a D5-brane with the worldvolume
$X^{123\circ}=0$.
We define the worldvolume coordinates $(x^0=t,x^4=x,x^a)$ ($a=5678$)
by the embedding
\begin{align}
X^\bullet&=\cosh x\cos t, & X^{5\sim 9}&=n^{5\sim 9}(x^a), \nonumber\\
X^0&=\cosh x\sin t, & X^\circ&=0, \nonumber\\
X^i&=0, \nonumber\\
X^4&=\sinh x,
\end{align}
where $n^m(x^a)$ is a five-dimensional unit vector parameterized by $x^a$.

To describe the worldvolume theory,
we need to introduce a local frame at each point $P$ on the
worldvolume.
This is achieved by giving an element $g(P)$ of the isometry group
$SO(2,4)\times SO(6)$ that takes the reference point $P_0$
to $P$,
and defining basis vectors $\bm{e}_M(P)$ at $P$ by
\begin{align}
\bm{e}_M(P)=g(P)\ul{\bm{e}}_M
\end{align}
where $\ul{\bm{e}}_M$ are the global basis vectors
associated with the global coordinates $X^M$ in the ambient space.
With this local frame, the Killing spinor $\varepsilon$
at a point $P$ is given by
\begin{align}
\varepsilon(P)=\rho(g^{-1})\xi,
\label{epsatp}
\end{align}
where $\rho$ is the representation
defined by (\ref{repxi1}) and (\ref{repxi2}).

In the following, 
we use local frames on the D5-brane specified by
\begin{align}
g=(e^{itT_0{}^\bullet}e^{ixT_4{}^\bullet})(g_{so(5)}(x^a)e^{\frac{\pi i}{2}T_9{}^\circ})
\label{localframe}
\end{align}
where $x^a$ are coordinates on $S^4$ and $g_{so(5)}(x^a)$ is an element of $so(5)_{56789}$ depending on $x^a$.
One can show that
the Killing spinor given by (\ref{epsatp}) with (\ref{localframe})
satisfies
\begin{align}
-\gamma^{045678}\sigma_x\varepsilon=\varepsilon
\label{d5susy}
\end{align}
at every point on the D5-brane worldvolume
provided that $\xi$ satisfies the relation (\ref{f04}).
This is nothing but the condition for the unbroken supersymmetry on the D5-brane.
Because the preserved supersymmetry is specified by the same condition (\ref{f04}) as for the fundamental line,
the fivebrane configuration is also $1/3$ BPS.

(\ref{d5susy}) relates the two components $\varepsilon_1$ and $\varepsilon_2$ of the Killing spinor,
and we can use $\varepsilon_1$ as the supersymmetry parameter
of the worldvolume theory.
By using (\ref{d5susy})
one can rewrite the Killing spinor equation as the differential equation
of $\varepsilon_1$
\begin{align}
D_\mu\varepsilon_1
&=\tfrac{1}{2}\gamma^{049}\gamma_\mu\sigma_z\varepsilon_1\quad(\mu=045678)
\label{killingond5}
\end{align}
On the D5-brane there are four scalar fields $\phi^i$ ($i=123$) and $\varphi$,
corresponding to fluctuations in $AdS_5$ and $S^5$, respectively,
as well as a fermion field $\lambda$ and a Maxwell field $a_\mu$ ($\mu=045678$).
We introduce the slash notation
\begin{align}
\sla D=\sum_{\mu=045678}\gamma^\mu D_\mu,\quad
\sla F=\frac{1}{2}\sum_{\mu,\nu=045678}\gamma^{\mu\nu}F_{\mu\nu},\quad
\sla\phi=\sum_{i=123}\gamma^i\phi^i,\quad
\sla\varphi=\gamma^9\varphi.
\end{align}
The action is
\begin{align}
{\cal L}&=-\tfrac{1}{4}f_{\mu\nu}f^{\mu\nu}
-\tfrac{1}{2}\partial_\mu\phi_i\partial^\mu\phi_i
-\tfrac{1}{2}\partial_\mu\varphi\partial^\mu\varphi
+4f_{04}\varphi
-\phi_i^2+2\varphi^2
\nonumber\\
&-\tfrac{1}{2}(\lambda\sla D\lambda)+\tfrac{1}{2}(\lambda\gamma^{049}\lambda).
\label{d5lag}
\end{align}
This is invariant under the transformation
\begin{align}
\delta\lambda&=\sla F\varepsilon_1+(\sla\phi-\sla\varphi)\gamma^{049}\varepsilon_1,\nonumber\\
\delta A_M&=(\lambda\gamma_M\varepsilon_1),
\end{align}
parameterized by the Killing spinor $\varepsilon_1$ satisfying (\ref{killingond5}).

\subsection{Modes on D5}\label{modesond5.app}

The symmetry on the D5-brane worldvolume is
$osp(4^*|4)$, and
the fluctuation modes belong to the representation
\begin{align}
{\cal R}_0\oplus
{\cal R}_1\oplus
{\cal R}_2\oplus\cdots
\end{align}
where each irreducible representation
${\cal R}_\ell$ consists of
modes in Table \ref{modes.tbl} \cite{Faraggi:2011bb,Faraggi:2011ge}.
\begin{table}[htb]
\caption{Modes belonging to the representation ${\cal R}_\ell$}\label{modes.tbl}
\centering
\begin{tabular}{cc|c|c|cc|c|c}
&                & $H$ & $J_1$ & $R_x$ & $R_y$ & Dynkin & range \\
\hline
[1]&
$\varphi,a_\mu$ & $\ell+3$ & $0$ & $\ell-1$ & $0$ & $[\ell-1,0]$ \\
{}[2]&
$\lambda$ & $\ell+\frac{5}{2}$ & $\frac{1}{2}$ & $\ell-\frac{1}{2}$ & $\frac{1}{2}$ & $[\ell-1,1]$ & $\ell=1,2,3,\ldots$ \\
{}[3]&
$a_m$ & $\ell+2$ & $0$ & $\ell$ & $1$ & $[\ell-1,2]$ \\
\hline
[4]&
$\phi^i$ & $\ell+2$ & $1$ & $\ell$ & $0$ & $[\ell,0]$ \\
{}[5]&
$\lambda$ & $\ell+\frac{3}{2}$ & $\frac{1}{2}$ & $\ell+\frac{1}{2}$ & $\frac{1}{2}$ & $[\ell,1]$ & $\ell=0,1,2,\ldots$ \\
{}[6]&
$\varphi,a_\mu$ & $\ell+1$ & $0$ & $\ell+1$ & $0$ & $[\ell+1,0]$ \\
\end{tabular}
\end{table}
The modes saturating the BPS bound,
as extracted from Table \ref{modes.tbl}, are shown in
Table \ref{bpsmodes.tbl}.
\begin{table}[htb]
\caption{BPS modes}\label{bpsmodes.tbl}
\centering
\begin{tabular}{llll}
modes & $(J_1,R_x,R_y)$ & range & index \\
\hline
modes in [5]
& $(\tfrac{1}{2},\ell+\tfrac{1}{2}-p,\tfrac{1}{2}+p)$ & $\ell\geq0$, $0\leq p\leq\ell$ & $-qx^{\ell-p}y^p$ \\
modes in [6]
& $(0,\ell+1-p,p)$ & $\ell\geq0$, $0\leq p\leq \ell+1$ & $x^{\ell+1-p}y^p$
\end{tabular}
\end{table}
Summing up all contributions, we obtain the letter index
of the fields on the D5-brane
\begin{align}
i_{\rm D5}
=\frac{1-q}{(1-x)(1-y)}-1
=\frac{x}{1-x}+\frac{y}{1-y}.
\end{align}

\paragraph{$U(1)_A$-refined index}

Because the orientifold preserves ${\cal N}=4$ supersymmetry,
all component fields in every ${\cal R}_\ell$ have the same orientifold parity
$(-1)^\ell$.
By the orientifolding,
modes with odd $\ell$ are projected out,
and the letter index is obtained by summing up the contributions from ${\cal R}_\ell$ with even $\ell$.
It is obtained from the refined index
\begin{align}
f_{\rm D5}=\frac{\eta-q}{(1-x)(1-y)}-\eta
\label{fd5refined2}
\end{align}
by applying the projection ${\cal P}_2$.

\section{Fat strings for $k\geq 3$}\label{fat.app}
\subsection{Boundary conditions on $(p,q)$-fivebrane junctions}

For the analysis of fluctuation modes on fat strings we need to know
boundary conditions on a fivebrane junction locus.
Because the boundary conditions are local conditions imposed at each point on
the junction locus, it is sufficient to consider a junction
consisting of flat semi-infinite fivebranes in the flat
spacetime.
The boundary conditions for scalar fields are
studied in \cite{Aharony:1997bh}.
See also \cite{Bergman:1998gs} for a similar analysis for $(p,q)$ string
junctions.
We derive the boundary conditions for the other fields
on the fivebranes so that they are consistent with supersymmetry.

We consider a $k$-pronged fivebrane junction consisting of $k$
semi-infinite fivebranes with charges $(p_j,q_j)$ ($j=1,\ldots,k$).
The fivebrane charges satisfy
\begin{align}
\sum_{j=1}^k p_j= \sum_{j=1}^k q_j =0.
\end{align}
In the derivation of the boundary conditions, we do not consider $S$-folding,
and $k$ semi-infinite fivebranes are treated as independent objects.
Let $\RR^{1,4}\times\ell_j$ be the worldvolume of the $j$-th fivebrane,
where $\RR^{1,4}$ is the Minkowski space along $01234$ directions
and $\ell_j$ is a semi-infinite line in the $89$ plane emanating from the
origin.
The common transverse directions are labeled by $567$.

We define some parameters for each fivebrane by
\begin{align}
(x^8+ix^9)|_{\ell_j}=ue^{i\theta_j},\quad
p_j+\tau q_j=t_ie^{i\theta'_j}.
\end{align}
where $\theta_j$ is the direction of the worldvolume of the $j$-th fivebrane in the $89$ plane,
and $u(\geq0)$ is the distance from the junction locus.
For each semi-infinite fivebrane we use $x^{0\sim 4}$ and $u$ as the worldvolume coordinates.
$\theta'_j$ is the argument of the central charge carried by the fivebrane,
and $t_j$ is the tension of the fivebrane normalized by the D5-brane tension.
$\theta_j$ and $\theta'_j$ are angular variables and are defined modulo $2\pi$.

The tension balance on the junction locus requires
\begin{align}
\sum_{j=1}^kt_je^{i\theta_j}=0
\label{tensionbalance}
\end{align}
and this is satisfied if $\theta_j-\theta'_j$ is $j$-independent;
\begin{align}
\theta_1-\theta_1'=
\theta_2-\theta_2'=\cdots=
\theta_k-\theta_k'.
\label{deltatheta}
\end{align}

One remark is in order.
(\ref{deltatheta}) contains $k-1$ constraints on the angles
and they cannot be obtained only from (\ref{tensionbalance}).
For a $k$-pronged junction with $k\geq 4$
(\ref{tensionbalance}) is not sufficient for stability
because even if
(\ref{tensionbalance}) holds
the junction may be unstable against
the splitting of the $k$-pronged junction locus
into multiple junction loci connected by
internal fivebranes.
To obtain a complete set of the stability conditions,
we slightly deform the $k$-pronged junction
into $k-2$ $3$-pronged junctions,
and impose the condition (\ref{tensionbalance})
on each of them.
Then we obtain
(\ref{deltatheta}).

If the condition (\ref{deltatheta}) holds,
the system is quarter BPS.
We take the coordinates $x^8$ and $x^9$ such that $\theta_j=\theta_j'$,
and then the supersymmetry projection condition
for each fivebrane is
\begin{align}
\gamma^{01234}(\cos\theta_j\gamma^8+\sin\theta_j\gamma^9)
(\cos\theta_j\sigma_x+\sin\theta_j\sigma_z)\xi=\xi,
\label{halfbps}
\end{align}
and all conditions are satisfied for $\xi$ satisfying
\begin{align}
\gamma^{012348}\sigma_x\xi=
\gamma^{89}(-i\sigma_y)\xi=\xi.
\end{align}

Let us consider fluctuation modes on a quarter BPS fivebrane junction.
On each semi-infinite fivebrane there are four scalar fields
associated with fluctuations of the brane.
Three of them, which are denoted by $\phi^{(j)}_m$ ($m=567$) are
fluctuation modes in the common transverse space, and the other,
denoted by $\varphi^{(j)}$, corresponds to the motion in the $89$ plane.
We also have the gauge field $a_\mu^{(j)}$ ($\mu=01234$) and fermion $\lambda^{(j)}$.

The transverse fluctuation $\phi_m^{(j)}$ must satisfy
the following boundary conditions
at $u=0$:
\begin{align}
\phi^{(1)}_m=\phi^{(2)}_m=\cdots =\phi^{(k)}_m,\quad
\sum_{j=1}^kt_j\partial_u\phi_m^{(j)}=0.
\label{phibc}
\end{align}
The first condition is necessary for the fivebranes to meet on the junction locus,
and the second condition comes from the tension balance along the transverse directions.

For the fields $\varphi$ representing the fluctuations in the $89$-plane
the following conditions are imposed.
\begin{align}
\partial_u\varphi^{(1)}=
\partial_u\varphi^{(2)}
=\cdots=\partial_u\varphi^{(k)},\quad
\sum_{j=1}^kt_j\varphi^{(j)}=0.
\label{varphibc}
\end{align}
The first condition is necessary to keep the stability conditions (\ref{deltatheta}),
and the second one is required by the geometric consistency.

Next, let us consider the gauge fields.
The boundary conditions imposed on the field strengths on the fivebranes
are
\begin{align}
f_{\mu u}^{(1)}=f_{\mu u}^{(2)}=\cdots=f_{\mu u}^{(k)},\quad
\sum_{j=1}^kt_jf_{\mu\nu}^{(j)}=0.
\label{gaugecond}
\end{align}
The $(p_j,q_j)$ fivebrane
with non-vanishing field strength $f_{\mu u}^{(j)}$
carries
$(p_j,q_j)$ string charges,
and the inflow of the string charges into the
junction is proportional to $(p_j,q_j)f_{\mu u}^{(j)}$.
The string charge conservation requires
$f_{\mu u}^{(1)}=f_{\mu u}^{(2)}=f_{\mu u}^{(3)}$ for a $3$-pronged junction.
For a $k$-pronged junction, we require such a condition to hold at every three-pronged junction obtained
by a deformation, and obtain the first relation in (\ref{gaugecond}).
The $(p_j,q_j)$ fivebrane with magnetic flux $f_{\mu\nu}^{(j)}$ carries D3 brane charge,
and the inflow is proportional to $t_if_{\mu\nu}^{(j)}$.
The D3-brane charge conservation requires the magnetic fluxes to satisfy
the second relation in
(\ref{gaugecond}).

The boundary conditions imposed on the fermion fields $\lambda^{(j)}$ are obtained
from those for bosonic fields by the supersymmetry.
To describe supersymmetry on each fivebrane,
we introduce
\begin{align}
\xi^{(j)}
=e^{\frac{\theta_j}{2}\gamma^{89}}e^{-i\frac{\theta'_j}{2}\sigma_y}\xi.
\end{align}
In terms of $\xi^{(j)}$,
the projection condition
(\ref{halfbps})
takes a $j$-independent form
\begin{align}
\gamma^{012348}\sigma_x\xi^{(j)}=\xi^{(j)}.
\end{align}

We use $\xi_1^{(j)}$ to describe supersymmetry on each fivebrane.
The Lagrangian density of the fields on the $j$-th fivebrane is
\begin{align}
{\cal L}^{(j)}=-\frac{1}{4}F^{(j)}_{MN}F^{(j)MN}-\frac{1}{2}(\lambda^{(j)}\sla\partial\lambda^{(j)}).
\end{align}
This is invariant
up to boundary terms
under the supersymmetry transformations
\begin{align}
\delta\lambda^{(j)}=\sla F^{(j)}\xi^{(j)}_1,\quad
\delta A_M^{(j)}=(\lambda^{(j)}\gamma_M\xi^{(j)}_1),
\label{susy10}
\end{align}
where we use ten-dimensional notation
\begin{align}
A_M^{(j)}=(a_\mu^{(j)},a_u^{(j)},\varphi^{(j)},\phi_m^{(j)}).
\end{align}

For each fivebrane, the parameter $\xi^{(j)}$ is divided into two parts
\begin{align}
\xi^{(j)}=\xi_+^{(j)}+\xi_-^{(j)},\quad
\gamma_R\sigma_z\xi_\pm^{(j)}=\pm\xi_\pm^{(j)},
\label{gammar}
\end{align}
where we defined
\begin{align}
\gamma_R=\gamma^{012349}
\end{align}
$\xi_-^{(j)}$ parameterizes the supersymmetry on the $j$-th fivebrane
which is broken by the other fivebranes.
The supersymmetry preserved by all the fivebranes is parameterized by
$\xi_+^{(j)}$.
$\xi_+^{(j)}$ is in fact independent of $j$, and we denote
the common value by $\xi_+$.

We also decompose the fermion $\lambda^{(j)}$ on each fivebrane into two parts by
\begin{align}
\lambda^{(j)}=\lambda^{(j)}_++\lambda^{(j)}_-,\quad
\gamma_R\lambda^{(j)}_\pm=\pm\lambda^{(j)}_\pm.
\end{align}
We require that the boundary conditions are invariant
under the supersymmetry transformation
parameterized by $\xi_{+,1}$.

The fermion transformation rule in 
(\ref{susy10}) splits into
\begin{align}
\delta\lambda^{(j)}_+=(\sla f^{(j)}_{\mu\nu}+\sla\partial_\mu\sla\varphi^{(j)}+\sla\partial_u\sla\phi_m^{(j)})\xi_{+,1},\quad
\delta\lambda^{(j)}_-=(\sla f^{(j)}_{\mu u}+\sla\partial_\mu\sla\phi_m^{(j)}+\sla\partial_u\sla\varphi^{(j)})\xi_{+,1}.
\end{align}
Consistency with these transformation rules requires the fermion fields to satisfy the boundary conditions
\begin{align}
\partial_u\lambda_+^{(1)}=\cdots=\partial_u\lambda_+^{(k)},\quad
\sum_{j=1}^k t_j\lambda^{(j)}_+=0,
\label{fermionbc1}
\end{align}
and
\begin{align}
\lambda_-^{(1)}=\cdots=\lambda_-^{(k)},\quad
\sum_{j=1}^k t_j\partial_u\lambda^{(j)}_-=0.
\label{fermionbc2}
\end{align}
(\ref{fermionbc1}) and
(\ref{fermionbc2}) are not independent but related by the
Dirac equation $\sla D\lambda=0$.

We obtained the boundary conditions
(\ref{phibc}),
(\ref{varphibc}),
(\ref{gaugecond}),
(\ref{fermionbc1}), and
(\ref{fermionbc2})
imposed on the fields on fivebranes
along the junction locus.
Now, let us derive the fluctuation modes
on a fat string with $k\geq 3$ using these boundary conditions.

\subsection{Mode analysis}
We assume that one of the $k$ fivebranes is a D5-brane
and its worldvolume is located at $\theta=0$.
The expectation value of the axiodilation field
is $\tau=\omega_k\equiv e^{\frac{2\pi i}{k}}$.
Then, the $j$-th fivebrane located at $\theta=\frac{2\pi j}{k}$
carries the complex charge $p+q\tau=\omega_k^j$,
and the tensions of all fivebranes are the same,
$t_j=|p+q\tau|=1$.

Let us define worldvolume coordinates
$x=X^5+iX^6$, $y=X^7+iX^8$, and $u=X^9$.
On the D5-brane
these are constrained by
\begin{align}
|x|^2+|y|^2+u^2=1,\quad
0\leq u\leq 1.
\end{align}
This is a $4$-hemisphere in $S^5$.

\subsubsection{Scalar fields $\phi_m$}
Because the fivebranes are not independent
and the $j$-th fivebranes ($j=1,\ldots,k-1$) is a $\ZZ_k$ image of the $k$-th fivebrane,
the scalar fields $\phi_m^{(j)}$ on different branes are related by
\begin{align}
\phi_m^{(j)}(x,y,u)=
\phi_m^{(k)}(\omega_k^jx,\omega_k^{-j}y,u).
\label{phimrel}
\end{align}
Let us expand $\phi_m^{(j)}$ into eigenfunctions of $R_1\equiv R_x-R_y$.
\begin{align}
\phi^{(j)}_m(x,y,u)=\sum_{R_1\in\ZZ}\phi^{(j)}_{m,R_1}(x,y,u)
\label{r1expansion}
\end{align}
Each eigenfunction satisfies
\begin{align}
\phi^{(j)}_{m,R_1}(\omega_kx,\omega_k^{-1}y,u)
=\omega_k^{R_1}\phi^{(j)}_{m,R_1}(x,y,u).
\label{r1eigenfunc}
\end{align}
Substituting
(\ref{r1expansion})
into the boundary conditions (\ref{phibc})
and using the relations
(\ref{phimrel}) and
(\ref{r1eigenfunc})
we obtain the following boundary conditions
for $R_1$ eigenmodes.
\begin{align}
(1-\omega^{R_1}_k)\phi^{(k)}_{m,R_1}=0,\quad
\sum_{j=1}^{k}\omega_k^{jR_1}
\partial_u\phi_{m,R_1}^{(k)}=0.
\end{align}
These conditions can be rewritten as follows.
\begin{align}
\left\{\begin{array}{ll}
\partial_u\phi^{(k)}_{m,R_1}=0 & (R_1=0\mod k)\\
\phi_{m,R_1}^{(k)}=0           & (R_1\neq 0\mod k)
\end{array}\right.
\end{align}
Namely, $\phi^{(k)}_{m,R_1}$ satisfies either the Dirichlet or the Neumann
boundary condition.
This implies that $\phi^{(k)}_{m,R_1}$ can be
extended to smooth functions on $S^4$ with
a specific reflection parity,
and can be expanded by $S^4$ harmonics.
The fields $\phi_{m,R_1}^{(k)}$ extended to $S^4$ satisfy
\begin{align}
{\cal R}\phi^{(k)}_{m,R_1}=s_k(R_1)\phi^{(k)}_{m,R_1},
\label{rphis}
\end{align}
where we defined the reflection operator
\begin{align}
{\cal R}\phi^{(k)}_{m,R_1}(x,y,u)=\phi^{(k)}_{m,R_1}(x,y,-u)
\end{align}
and the sign function
\begin{align}
s_k(R_1)=
\left\{\begin{array}{ll}
+1 & (R_1=0\mod k)\\
-1 & (R_1\neq 0\mod k)
\end{array}\right.
\end{align}

As we mentioned above, $\phi_m$ can be expanded by $S^4$ harmonics.
The equation of motion for $\phi_m$ is
\begin{align}
(D_{AdS_2}^2+D_{S^4}^2-2)\phi_m=0\quad
(m=123)
\end{align}
We expand $\phi_{m,R_1}^{(k)}$ in terms of the scalar $S^4$ harmonics
$Y^{(\ell,0)}_{R_1}$ and scalar AdS harmonics ${\cal Y}^{(E)}$ by
\begin{align}
\phi_{m,R_1}^{(k)}=\sum(\text{coeff}){\cal Y}^{(E)}Y_{R_1}^{(\ell,0)}
\end{align}
See Appendix \ref{harmonics.sec} for definitions and formulas of the harmonics.
The harmonics satisfy
\begin{align}
D_{AdS_2}^2{\cal Y}^{(E)}=E(E-1){\cal Y}^{(E)},\quad
D_{S^4}^2Y^{(\ell,0)}=-\ell(\ell+3)Y^{(\ell,0)},
\label{dscalar}
\end{align}
and we obtain the relation
\begin{align}
E(E-1)-\ell(\ell+3)-2=0.
\end{align}
There are two solutions:
$E=\ell+2$ and
$E=-\ell-1$,
and the positive frequency solution $E=\ell+2$ corresponds to mode [4] in Table \ref{modes.tbl}.

The $S^4$ harmonics satisfy
\begin{align}
{\cal R}Y^{(\ell,0)}_{R_1}
=(-1)^{\ell+R_1}Y^{(\ell,0)}_{R_1}
\label{scalarry}
\end{align}

Combining (\ref{rphis}) and (\ref{scalarry}) we obtain the selection rule
\begin{align}
(-1)^{\ell+R_1}=s_k(R_1).
\label{selection}
\end{align}
In the following
we will explicitly show that the same selection rule applies
to all the other fields, too.

\subsubsection{$\varphi$ and $f_{04}$}
Due to the term $4\varphi f_{04}$ in the Lagrangian (\ref{d5lag}),
the scalar field $\varphi$ is mixed with the gauge field.
By adopting the gauge fixing condition $D_ma^m=0$ ($m=5678$)
the equations of motion
are closed in two scalar fields $\varphi$ and $f\equiv f_{04}$.
\begin{align}
&D_{AdS_2}^2f+D_{S^4}^2f+4D_{AdS_2}^2\varphi=0,
\nonumber\\
&D_{AdS_2}^2\varphi+D_{S^4}^2\varphi+4f+4\varphi=0.
\label{fphieom}
\end{align}
Let us expand $f$ and $\varphi$ by scalar harmonics:
\begin{align}
f=\alpha{\cal Y}^{(E)}Y^{(\ell,0)},\quad
\varphi=\beta{\cal Y}^{(E)}Y^{(\ell,0)}.
\end{align}
Substituting these into the equations
of motion (\ref{fphieom}) and using the properties of
the scalar harmonics (\ref{dscalar}),
we obtain the following equation for the coefficients.
\begin{align}
\left(\begin{array}{cc}
E(E-1)-\ell(\ell+3) & 4E(E-1) \\
4 & E(E-1)-\ell(\ell+3)+4 \\
\end{array}\right)
\left(\begin{array}{c}
\alpha \\ \beta
\end{array}\right)
=0.
\label{matrixab}
\end{align}
The determinant of the matrix in
(\ref{matrixab}) is
\begin{align}
\det=(E-\ell)(E-\ell-4)(E+\ell-1)(E+\ell+3),
\end{align}
and we obtain four values of $E$ for each $\ell$.
Two of them, $E=\ell$ and $E=\ell+4$ are positive frequencies,
and the corresponding modes belong to ${\cal R}_{\ell-1}$ and ${\cal R}_{\ell+1}$,
respectively.
We need to shift $\ell$ by $\pm1$ to obtain modes in ${\cal R}_\ell$.
Namely, the modes $[6]$ and $[1]$ in Table \ref{modes.tbl} are
\begin{align}
(f,\varphi)\sim{\cal Y}^{(\ell+1)}Y^{(\ell+1,0)},\quad
{\cal Y}^{(\ell+3)}Y^{(\ell-1,0)}.
\label{aphimodes}
\end{align}

The boundary conditions satisfied by $\varphi$ and $f_{04}$ are
the same as those for $\partial_u\phi_m$.
Therefore, the conditions for the quantum numbers are
opposite to those for $\phi_m$ due to the minus sign from $\partial_u$:
\begin{align}
{\cal R}\varphi^{(k)}_{R_1}=-s_k(R_1)\varphi^{(k)}_{R_1},\quad
{\cal R}f^{(k)}_{04,R_1}=-s_k(R_1)f^{(k)}_{04,R_1}.
\label{varphifbc}
\end{align}
Using the property (\ref{scalarry}) of the $S^4$ scalar harmonics,
we can show that the modes in
(\ref{aphimodes}) satisfy
\begin{align}
{\cal R}\varphi=-(-1)^{\ell+R_1}\varphi,\quad
{\cal R}f_{04}=-(-1)^{\ell+R_1}f_{04}.
\end{align}
Combining this with
(\ref{varphifbc})
we obtain the selection rule (\ref{selection}).

\subsubsection{$S^4$ vector}
If we adopt the gauge fixing condition $D_\alpha a^\alpha=0$ ($\alpha=04$),
the Maxwell equation containing the $S^4$ vector field $a_m$ becomes
\begin{align}
D_{AdS_2}^2a^m+D_nf^{nm}=0.
\end{align}
We expand the field in terms of the scalar $AdS_2$ harmonics ${\cal Y}^{(E)}$
and vector $S^4$ harmonics $Y_m^{(\ell,1)}$ by
\begin{align}
a_m={\cal Y}^{(E)}Y_m^{(\ell,1)}.
\end{align}
The vector harmonics satisfy
\begin{align}
D^n(D_nY_m^{(\ell,1)}-D_mY_n^{(\ell,1)})=-(\ell+1)(\ell+2)Y_m^{(\ell,1)}.
\end{align}
Using this and the property of the scalar harmonics in (\ref{dscalar}),
we obtain
\begin{align}
E(E-1)-(\ell+1)(\ell+2)=0
\end{align}
and the positive frequency is
\begin{align}
E=\ell+2.
\end{align}
This corresponds to the mode [3] in Table \ref{modes.tbl}.

The components $f_{mn}^{(i)}$ and $f_{mu}^{(i)}$ of the gauge field strength
satisfy the same boundary condition as $\partial_u\phi_a$ and $\phi_a$, respectively.
Therefore, the conditions for the modes satisfying the boundary conditions are
\begin{align}
f^{(k)}_{mn,R_1}(-u)=-s_k(R_1)f^{(k)}_{mn,R_1}(u),\quad
f^{(k)}_{mu,R_1}(-u)=+s_k(R_1)f^{(k)}_{mu,R_1}(u).
\end{align}
These can be combined into
\begin{align}
{\cal R}(f^{(k)}_{mn,R_1},f^{(k)}_{mu,R_1})
=-s_k(R_1)(f^{(k)}_{mn,R_1},f^{(k)}_{mu,R_1})
\label{rf0}
\end{align}
where the reflection acts on the components by
\begin{align}
{\cal R}f^{(k)}_{mn}(u)
={\cal R}f^{(k)}_{mn}(-u),\quad
{\cal R}f^{(k)}_{mu}(u)
=-{\cal R}f^{(k)}_{mu}(-u).
\end{align}

The $2$-form harmonics in $S^4$ satisfy
\begin{align}
{\cal R}Y^{(\ell,1)}_{R_1}=(-1)^{\ell+R_1+1}Y^{(\ell,1)}_{R_1}
\label{refvec}
\end{align}

Combining 
(\ref{rf0}) and
(\ref{refvec}) we obtain
the selection rule (\ref{selection}).

\subsubsection{Fermion field $\lambda$}
The Dirac equation for $\lambda$ is
\begin{align}
\sla D\lambda-\gamma^{049}\lambda=0.
\label{diraclambda}
\end{align}
In order to solve this, we adopt the following
representation of the Dirac matrices.
\begin{align}
\gamma^\alpha&=\gamma_{(2)}^\alpha\otimes 1_4\otimes 1_2\otimes\sigma_x,(\alpha=04)\nonumber\\
\gamma^m&=\ol\gamma_{(2)}\otimes\gamma_{(4)}^m\otimes1_2\otimes\sigma_x,(m=5678)\nonumber\\
\gamma^9&=\ol\gamma_{(2)}\otimes\ol\gamma_{(4)}\otimes1_2\otimes\sigma_x,\nonumber\\
\gamma^i&=1_2\otimes1_4\otimes\gamma_{(3)}^i\otimes\sigma_y,(i=123)\nonumber\\
\ol\gamma&=1_2\otimes1_4\otimes1_2\otimes\sigma_z,
\label{gammareps}
\end{align}
where $\gamma_{(2)}^\alpha$ and $\ol\gamma_{(2)}$ are the two-dimensional
Dirac matrices and the corresponding chirality matrix
\begin{align}
\gamma_{(2)}^0=i\sigma_y,\quad
\gamma_{(2)}^4=i\sigma_x,\quad
\ol\gamma_{(2)}=i\sigma_z,
\end{align}
and $\gamma_{(4)}^m$ are the four-dimensional
Dirac matrices, and $\ol\gamma_{(4)}$ is the corresponding chirality matrix
\begin{align}
\gamma_{(4)}^5=\sigma_x\otimes\sigma_x,\quad
\gamma_{(4)}^6=\sigma_x\otimes\sigma_y,\quad
\gamma_{(4)}^7=\sigma_x\otimes\sigma_z,\quad
\gamma_{(4)}^8=\sigma_y\otimes1_2,\quad
\ol\gamma_{(4)}=\sigma_z\otimes1_2.
\end{align}
Then, the Dirac operator (including the non-derivative term) becomes
\begin{align}
\sla D-\gamma^{049}=
[(\gamma_{(2)}^\alpha\otimes 1)D_\alpha+(\ol\gamma_{(4)}\otimes\gamma^m)D_m-(1_2\otimes\ol\gamma_{(4)})]
\otimes1_2\otimes\sigma_x
\end{align}
We expand the fermion field as follows.
\begin{align}
\lambda=
(\alpha {\cal Y}_L^{(E)}Y_L^{(\ell+\frac{1}{2},\frac{1}{2})}+
\beta {\cal Y}_L^{(E)}Y_R^{(\ell+\frac{1}{2},\frac{1}{2})}+
\gamma {\cal Y}_R^{(E)}Y_L^{(\ell+\frac{1}{2},\frac{1}{2})}+
\delta {\cal Y}_R^{(E)}Y_R^{(\ell+\frac{1}{2},\frac{1}{2})})
\otimes\eta\otimes \uparrow,
\end{align}
where 
$Y_{L/R}^{(\ell+\frac{1}{2},\frac{1}{2})}$ are spinor harmonics
satisfying the relations
\begin{align}
\left\{\begin{array}{l}
\gamma_{(2)}^\alpha D_\alpha {\cal Y}^{(E)}_L=-i(E-\frac{1}{2}){\cal Y}_R^{(E)}\\
\gamma_{(2)}^\alpha D_\alpha {\cal Y}^{(E)}_R=+i(E-\frac{1}{2}){\cal Y}_L^{(E)}
\end{array}\right.
\quad
\left\{\begin{array}{l}
\gamma_{(4)}^m D_m Y^{(\ell+\frac{1}{2},\frac{1}{2})}_L=+(\ell+2)Y^{(\ell+\frac{1}{2},\frac{1}{2})}_R\\
\gamma_{(4)}^m D_m Y^{(\ell+\frac{1}{2},\frac{1}{2})}_R=-(\ell+2)Y^{(\ell+\frac{1}{2},\frac{1}{2})}_L
\end{array}\right.
\end{align}
Under the reflection
${\cal R}_{(4)}f(u)=\gamma_8f(-u)$
the spinor harmonics are transformed by
\begin{align}
{\cal R}_{(4)}Y_{L,R_1}^{(\ell+\frac{1}{2},\frac{1}{2})}=(-1)^{\ell+R_1}Y_{R,R_1}^{(\ell+\frac{1}{2},\frac{1}{2})},\quad
{\cal R}_{(4)}Y_{R,R_1}^{(\ell+\frac{1}{2},\frac{1}{2})}=(-1)^{\ell+R_1}Y_{L,R_1}^{(\ell+\frac{1}{2},\frac{1}{2})},
\label{ref4}
\end{align}

Substituting these into the Dirac equation (\ref{diraclambda}),
we obtain the matrix equation for the coefficient vector.
\begin{align}
\left(\begin{array}{cccc}
-1 & \ell+2 & -i(E-\frac{1}{2}) & 0 \\
-(\ell+2) & 1 & 0 & -i(E-\frac{1}{2}) \\
i(E-\frac{1}{2}) & 0 & -1 & -(\ell+2) \\
0 & i(E-\frac{1}{2}) & \ell+2 & 1
\end{array}\right)
\left(\begin{array}{c}
\alpha \\ \beta \\ \gamma \\ \delta
\end{array}\right)=0.
\label{matrixeq}
\end{align}
The determinant of the matrix in (\ref{matrixeq}) is
$\prod_{i=1}^4(E-E_i)$ with
\begin{align}
E_1=-(\ell+\tfrac{5}{2}),\quad
E_2=-(\ell+\tfrac{1}{2}),\quad
E_3=\ell+\tfrac{7}{2},\quad
E_4=\ell+\tfrac{3}{2},
\end{align}
and the corresponding eigenvectors are
\begin{align}
\left(\begin{array}{c}
\alpha \\ \beta \\ \gamma \\ \delta
\end{array}\right)
=
\left(\begin{array}{c}
i \\ -i \\ 1 \\ 1
\end{array}\right),
\left(\begin{array}{c}
i \\ i \\ -1 \\ 1
\end{array}\right),
\left(\begin{array}{c}
-i \\ i \\ 1 \\ 1
\end{array}\right),
\left(\begin{array}{c}
-i \\ -i \\ -1 \\ 1
\end{array}\right),
\end{align}
respectively.

We are interested in the positive frequency modes with
\begin{align}
E=\ell+\tfrac{3}{2},\quad\ell+\tfrac{7}{2}
\end{align}
The mode with $E=\ell+\frac{3}{2}$, which corresponds to [5] in Table \ref{modes.tbl},
is given by
\begin{align}
\lambda=-i{\cal Y}_L^{(\ell+\frac{3}{2})}(Y_L^{(\ell+\frac{1}{2},\frac{1}{2})}+Y_R^{(\ell+\frac{1}{2},\frac{1}{2})})-{\cal Y}_R^{(\ell+\frac{3}{2})}(Y_L^{(\ell+\frac{1}{2},\frac{1}{2})}-Y_R^{(\ell+\frac{1}{2},\frac{1}{2})}).
\label{mode2}
\end{align}
The other positive frequency mode with eigenvalue $E=\ell+\frac{7}{2}$
belongs not to ${\cal R}_\ell$ but to ${\cal R}_{\ell+1}$.
We need to shift $\ell$ by one to obtain the mode in ${\cal R}_\ell$.
Namely,
the mode [2] in Table \ref{modes.tbl} is given by
\begin{align}
\lambda=-i{\cal Y}_L^{(\ell+\frac{5}{2})}(Y_L^{(\ell-\frac{1}{2},\frac{1}{2})}-Y_R^{(\ell-\frac{1}{2},\frac{1}{2})})
-{\cal Y}_R^{(\ell+\frac{5}{2})}(Y_L^{(\ell-\frac{1}{2},\frac{1}{2})}+Y_R^{(\ell-\frac{1}{2},\frac{1}{2})}).
\label{mode5}
\end{align}

Let us impose the boundary conditions
on (\ref{mode2}) and (\ref{mode5}).
As shown in (\ref{fermionbc1}) and (\ref{fermionbc2}),
$\lambda_{-,R_1}^{(j)}$
and $\lambda_{+,R_1}^{(j)}$ satisfy the same boundary conditions
as $\phi_m$ and $\partial_u\phi_m$, respectively.
This means the $R_1$ eigenmodes satisfy the conditions
\begin{align}
\lambda_{-,R_1}^{(k)}(-u)=+s_k(R_1)\lambda_{-,R_1}^{(k)}(u),\quad
\lambda_{+,R_1}^{(k)}(-u)=-s_k(R_1)\lambda_{+,R_1}^{(k)}(u).
\end{align}
These can be combined into
\begin{align}
\gamma_R\lambda_{R_1}^{(k)}(-u)
=-s_k(R_1)\lambda_{R_1}^{(k)}(u).
\label{grlambda}
\end{align}
With the representation (\ref{gammareps}),
the left hand side of (\ref{grlambda}) is rewritten as
\begin{align}
\gamma_R\lambda_{R_1}^{(k)}(-u)=
(\ol\gamma_{(2)}\otimes 1\otimes1\otimes 1)
{\cal R}_{S^4}\lambda_{R_1}^{(k)}(-u)
\end{align}
where ${\cal R}_{S^4}$ is the reflection operator acting on
spinors in $S^4$:
\begin{align}
{\cal R}_{S^4}\psi(u)
=\gamma_{(4)}^8\psi(-u)
\label{rlambda1}
\end{align}
Using (\ref{ref4}) we obtain
\begin{align}
{\cal R}\lambda_{R_1}=-(-1)^{\ell+R_1}\lambda_{R_1}
\label{reflambda}
\end{align}
for both
(\ref{mode2}) and (\ref{mode5}).
Combining this with
(\ref{grlambda}) we obtain the
selection rule (\ref{selection}).

\subsection{Superconformal index}
Let us calculate the letter index of BPS modes on the fat string.
The BPS modes are contained in [5] and [6] in Table \ref{modes.tbl},
and the following relation holds for them.
\begin{align}
\ell=H-J_1-1.
\label{ellhj1}
\end{align}

We can rewrite the left hand side of the selection rule (\ref{selection}) as
\begin{align}
(-1)^{\ell+R_1}=(-1)^{H-J_1+R-1}
=(-1)^{1+2R_x}=(-1)^{F+1},
\end{align}
where $F$ is the fermion number.
The selection rule (\ref{selection}) for BPS states becomes
\begin{align}
(-1)^{F+1}=s_k(R_1).
\end{align}
Namely, we take only the fermionic (bosonic) contribution
for $R_1=0\mod k$ ($R_1\neq0\mod k$).
We obtain the letter index
\begin{align}
i_k=
\frac{1}{1-q}\left(\frac{x}{1-x}+\frac{y}{1-y}+1\right)
-\frac{1+q}{1-q}\left(\frac{x^k}{1-x^k}+\frac{y^k}{1-y^k}+1\right)
\end{align}
for the fat string.

\subsection{$S^4$ and $AdS_2$ harmonics}\label{harmonics.sec}
\subsubsection{$S^4$ harmonics}
Let us consider spin-$s$ $S^4$ harmonics
belonging to an $so(5)$ irreducible representation $r$.
There are $\dim r$ linearly independent harmonics
and we collectively denote them by $Y^r$.
In the main text we are interested in
harmonics with a specific value of $R_1\equiv R_x-R_y$,
and such harmonics are collectively denoted by $Y^r_{R_1}$.
We use the highest weight $(R_x^h,R_y^h)$ to specify the representation $r$.
To keep the notation concise, we omit
the spin when it is clear from the context.

\paragraph{Scalar harmonics}
The $S^4$ scalar harmonics
belong to the irreducible representations
$(\ell,0)$ ($\ell=0,1,\ldots$)
and we denote them by $Y^{(\ell,0)}$.
One can realize each of them as the restriction to $S^4$
of a homogeneous scalar harmonic function of degree $\ell$
in the ambient space $\RR^5$.
Let $(x,y,u)$ be the coordinates in the ambient space.
$x$ and $y$ are complex variables rotated by $R_x$ and $R_y$ respectively,
and $u$ is a real coordinate.
From the harmonic condition in the ambient space one can easily obtain
\begin{align}
D_{S^4}^2Y^{(\ell,0)}=-\ell(\ell+3)Y^{(\ell,0)}
\end{align}
Each of the harmonics $Y^{(\ell,0)}$ is a linear combination of
terms of the form
\begin{align}
x^{n_x}\bar x^{n_{\bar x}}y^{n_y}\bar y^{n_{\bar y}}u^{n_u},\quad(n_x+n_{\bar x}+n_y+n_{\bar y}+n_u=\ell),
\label{scalarhterms}
\end{align}
and then it satisfies
\begin{align}
Y^{(\ell,0)}(-x,-y,-u)=
(-1)^\ell Y^{(\ell,0)}(x,y,u).
\end{align}
The harmonics $Y^{(\ell,0)}_{R_1}$ with the value of $R_1$ specified
are linear combinations of terms
(\ref{scalarhterms}) with $n_x-n_{\bar x}-n_y+n_{\bar y}=R_1$.
Then, they satisfy
\begin{align}
Y_{R_1}^{(\ell,0)}(x,y,-u)
=(-1)^{\ell+R_1}Y_{R_1}^{(\ell,0)}(x,y,u).
\end{align}

\paragraph{$1$-form and $2$-form harmonics}
The Maxwell potential field and the field strength in $S^4$ are expanded
by $1$-form and $2$-form harmonics.
$1$ form harmonics belong to the representations $(\ell,0)$ or $(\ell,1)$,
and the former are associated with gauge degrees of freedom, and we are interested
in the latter.

Let $Y^{(\ell,1)}_1=Y^{(\ell,1)}_mdx^m$ be $1$-form harmonics, and $Y_2^{(\ell,1)}=dY_1^{(\ell,1)}=\frac{1}{2}Y_{mn}^{(\ell,1)}dx^mdx^n$ be the
corresponding $2$-form harmonics.
These are realized as the restriction to $S^4$ of a Maxwell field in the ambient space
which is homogeneous and 
satisfies the five-dimensional Maxwell equations.
From the five-dimensional Maxwell equations one can obtain
\begin{align}
D_{S^4}^mY_{mn}^{(\ell,1)}=-(\ell+1)(\ell+2)Y_n^{(\ell,1)}.
\end{align}
Each two-form harmonic $Y_2^{(\ell,1)}$ is the pull-back
of the five-dimensional two-form field strength $F_2$, which
is a linear combination of
terms in the form
\begin{align}
F_2\sim x^{n_x-1}\ol x^{n_{\ol x}}
y^{n_y-1}\ol y^{n_{\ol y}}
u^{n_u}
dxdy,\quad
(n_x+n_{\ol x}+n_y+n_{\ol y}+n_u=\ell+1)
\label{twoformterms}
\end{align}
and other terms with other differentials $dxd\ol y$, $dxdu$, and so on.
Each $n_i$ counts the number of the variable in the term including the differential part.

Two-form harmonics $Y_{2,R_1}^{(\ell,1)}$ with a specific value of $R_1$ are
restrictions of linear combinations of terms
(\ref{twoformterms}) with $n_x-n_{\bar x}-n_y+n_{\bar y}=R_1$,
and their reflection parity is
\begin{align}
{\cal R}Y^{(\ell,1)}_{2,R_1}=(-1)^{\ell+R_1+1}Y^{(\ell,1)}_{2,R_1},
\end{align}
where the reflection ${\cal R}$ flips the signs of both $u$ and $du$.

\paragraph{Spinor harmonics}

Spinor harmonics in $S^4$ belong to the $so(5)$ representations
$(\ell+\frac{1}{2},\frac{1}{2})$ ($\ell=0,1,\ldots$).
There are two spinor representations with different chiralities,
and we denote spinor harmonics with positive and negative chirality
by $Y^{(\ell+\frac{1}{2},\frac{1}{2})}_L$ and $Y^{(\ell+\frac{1}{2},\frac{1}{2})}_R$, respectively.
They can be given as the restriction of a five-dimensional
spinor field satisfying the massless Dirac equation.
From a single spinor function $\Psi$ in the ambient space, we obtain
two spinor harmonics with opposite chiralities.
Let
$Y_L^{(\ell+\frac{1}{2},\frac{1}{2})}$
and $Y_R^{(\ell+\frac{1}{2},\frac{1}{2})}$ denote such
a pair of harmonics.
From the five-dimensional Dirac equation, we obtain
\begin{align}
\left\{\begin{array}{l}
\gamma_{(4)}^m D_m Y^{(\ell+\frac{1}{2},\frac{1}{2})}_L=+(\ell+2)Y^{(\ell+\frac{1}{2},\frac{1}{2})}_R\\
\gamma_{(4)}^m D_m Y^{(\ell+\frac{1}{2},\frac{1}{2})}_R=-(\ell+2)Y^{(\ell+\frac{1}{2},\frac{1}{2})}_L
\end{array}\right.
\label{diracy}
\end{align}
Note that the Dirac operator $\sla D_{S^4}$
flips the chirality and 
(\ref{diracy}) gives the relations between harmonics with opposite chiralities.
The five dimensional spinor field belonging to the $(\ell+\frac{1}{2},\frac{1}{2})$
representation is a linear combination of terms like
\begin{align}
\Psi\sim x^{n_x}\ol x^{n_{\ol x}}
y^{n_y}\ol y^{n_{\ol y}}
u^{n_u}
\chi_{s_xs_y},\quad
(n_x+n_{\ol x}+n_y+n_{\ol y}+n_u=\ell)
\label{psiterms}
\end{align}
where $\chi_{s_xs_y}$ ($s_x,s_y=\pm\frac{1}{2}$)
is a constant spinor in the ambient space carrying $R_x=s_x$ and $R_y=s_y$.

The reflection of the spinor function $\Psi$ is given by%
\footnote{On the reflection plane the local basis $\bm{e}_8$
is identified with
the global basis $-\bm{e}_{9}$.}
\begin{align}
{\cal R}\Psi(x,y,u)
&=-\gamma_{(5)}^{9}\Psi(x,y,-u)
\nonumber\\
&=(-i\gamma_{(5)}^{56})(-i\gamma_{(5)}^{78})\Psi(x,y,-u)
\nonumber\\
&=(2s_x)(2s_y)\Psi(x,y,-u)
\nonumber\\
&=(-1)^{s_x-s_y+n_u}\Psi(x,y,u),
\end{align}
where we used $\gamma_{(5)}^{56789}=1$ at the second equality,
and the spin matrices acting on $\Psi$ are $s_x=-\frac{i}{2}\gamma^{56}$ and $s_y=-\frac{i}{2}\gamma^{78}$
at the third equality.
For a five-dimensional spinor function $\Psi_{R_1}$ with a fixed value of $R_x$,
the exponents in (\ref{psiterms}) are constrained by
\begin{align}
R_1
=n_x-n_{\bar x}-n_y+n_{\bar y}+s_x-s_y,
\end{align}
and the reflection parity is
\begin{align}
{\cal R}\Psi_{R_1}
&=(-1)^{R_1+\ell}\Psi_{R_1}.
\end{align}
The corresponding relations for the spinor harmonics are
\begin{align}
{\cal R}_{(4)}Y_{L,R_1}^{(\ell+\frac{1}{2},\frac{1}{2})}=(-1)^{\ell+R_1}Y_{R,R_1}^{(\ell+\frac{1}{2},\frac{1}{2})},\quad
{\cal R}_{(4)}Y_{R,R_1}^{(\ell+\frac{1}{2},\frac{1}{2})}=(-1)^{\ell+R_1}Y_{L,R_1}^{(\ell+\frac{1}{2},\frac{1}{2})}
\end{align}

\subsubsection{$AdS_2$ harmonics}

$AdS_2$ harmonics belong to $so(1,2)\sim sl(2)$ representations.
Let $L_0$ and $L_{\pm1}$ be the generators defined in the standard way.
The highest weight of
an irreducible representation can be specified by
the energy of the lowest energy mode,
and other descendant modes are obtained by applying the
raising operator $L_{-1}$.
Let us collectively denote scalar harmonics by ${\cal Y}^{(E)}$.
The following relations hold.
\begin{align}
D_{AdS_2}^2{\cal Y}^{(E)}=E(E-1){\cal Y}^{(E)}.
\end{align}
We denote spinor harmonics with positive and negative chiralities
by ${\cal Y}_L^{(E)}$ and ${\cal Y}_R^{(E)}$, respectively.
The following relations hold:
\begin{align}
\sla D_{AdS_2}{\cal Y}_L^{(E)}=-i(E-\tfrac{1}{2}){\cal Y}_R^{(E)},\quad
\sla D_{AdS_2}{\cal Y}_R^{(E)}=+i(E-\tfrac{1}{2}){\cal Y}_L^{(E)}.
\end{align}


\end{document}